\documentclass[%
 reprint,
 amsmath,amssymb,
 aps,
]{revtex4-2}

\usepackage{graphicx}
\usepackage{dcolumn,multirow}
\usepackage{bm}

\begin{document}


\title{Paraxial forward-scatter field analysis for a THz pulse traveling down \\ a highly overmoded iris-line waveguide}

\author{Adham Naji}\email{anaji@stanford.edu}
 \altaffiliation[]{}
\author{Gennady Stupakov}
 
\affiliation{%
 SLAC National Linear Accelerator Laboratory, Stanford University, Menlo Park, CA 94025
}%

\date{\today}


\begin{abstract}
Highly overmoded iris-line structures carry desirable features for the efficient transportation of THz radiation over long distances. Previous studies have analyzed the iris line, modeled approximately as a long open-resonator structure with thin screens, using methods such as Vainstein's impedance boundary condition or perturbative mode matching. The aim in those methods was to seek the eigensolution that represents the dominant (least lossy) propagation mode in a long iris-line structure at steady state. In this paper, a forward-traveling wave analysis is presented, wherein a short THz pulse paraxially traverses the oversized structure cell by cell, including the transient regime. The iris line's periodic discontinuities are analysed in terms of forward-wave orthogonal mode decompositions, which are then used to build a `forward-scatter' model (matrix) for each cell.  The presented approach predicts a diffraction loss behavior that agrees with that of the eigensolution found by the previous analytical methods at the steady state for a long waveguide, but adds the ability of analyzing waveguides of finite lengths, showing transients as they evolve and settle to the dominant mode down the line, and offers numerical implementation speeds that are typically an order-of-magnitude faster than those previously reported for the mode-matching method. This approach may also be easily extended to analyze iris lines with mechanical imperfections (e.g.~misalignments or aperiodic defects), allowing for investigations of performance sensitivity against manufacturing tolerances. 
\end{abstract}

\maketitle



\section{\label{Intro}Introduction}

\begin{figure*}
\includegraphics[width=0.9\textwidth]{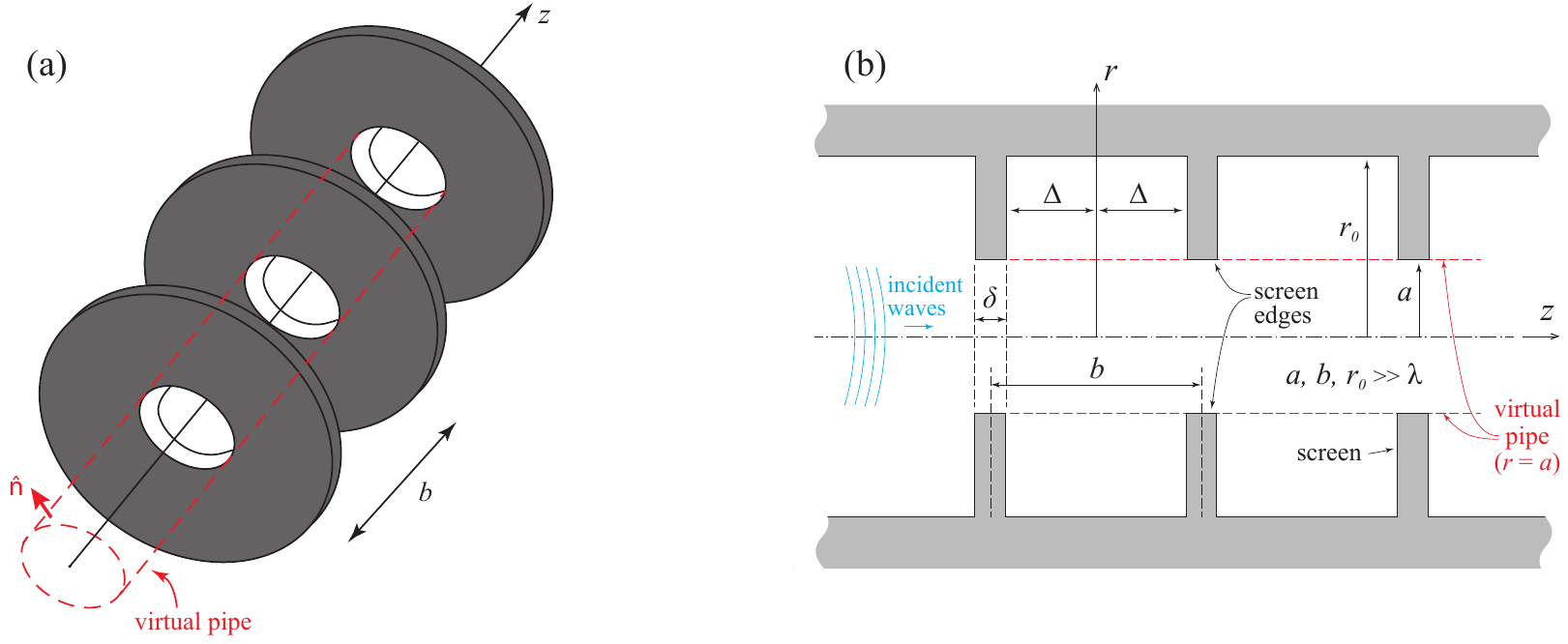}
\caption{\label{fig:geometry}  (a) The iris-line geometry in three-dimensions, showing the screens without the enclosing chamber at the outer radius. (b) A cross section in the iris line, showing the enclosing chamber.}
\end{figure*}

The highly overmoded iris-line structure (Fig.~\ref{fig:geometry}) has been recently investigated as a candidate solution for the efficient transportation of THz radiation over long distances~\cite{DESY, NajiTHz1}. One application example, which was discussed in references \cite{DESY, NajiTHz1}, is the transport of radiation in the 3--15 THz range (20--100~$\mu$m wavelength), generated by a THz linearly-polarized undulator, over hundreds of meters at the Linac Coherent Light Source (LCLS), SLAC National Accelerator Laboratory. Advantages offered by the iris line compared to other transport methods, such as beam relaying using a combination of planar and paraboloidal mirrors \cite{NajiTHz1,Zhang,DESY2}, include its lower propagation loss over long distance (e.g.~around 10--15\% power loss, compared to approximately 30\%, for 150~m distance \cite{NajiTHz1,Zhang}) and its simple geometrical setup. Specifically, the power loss for the dominant dipole mode on the iris line is well-approximated in the limit of thin screens by the following law, derived from Vainstein's impedance boundary condition \cite{DESY,Vainstein1,Vainstein2},
\begin{equation}
    L\approx \left[1-e^{-4.75 c^{3/2} b^{1/2} \omega^{-3/2}a^{-3}z} \right]\times 100\%, \label{VainsteinLoss__}
\end{equation}
where $c$ is the speed of light in vacuum, $\omega$ is the angular frequency, $a$ is the radius of the iris, $b$ is the period of the iris line and $z$ is the distance travelled down the line. The attenuation constant (exponent) is seen to drop with larger radii and higher frequencies (proportional to $a^{-3}$ and $\omega^{-3/2}$).  This mode has also been shown to preserve the linear polarization across the iris and support an amplitude profile that follows the $J_{0}(2.4 r/a)$ Bessel function profile, which happens to overlap conveniently with the approximate Gaussian profile typically emitted from a linear undulator \cite{NajiTHz1}, where $r$ denotes the radial distance from the axis $z$.  Although the law (\ref{VainsteinLoss__}) is a good approximation in general, when the thin screens have finite (nonzero) thickness, $\delta > 0$, this law starts to overestimate diffraction loss and underestimate ohmic loss at the screen edges. The predictions of this law can be corrected, or superseded, by the estimates found from the mode-matching technique \cite{NajiTHz1}, but at a higher computational price. As $\delta$ increases, diffraction loss is decreased (as the gaps between the screens start to gradually close up) and ohmic loss at the screen edges starts to increase \cite{NajiTHz1}. For relatively thin screens ($\delta\ll b$), however, ohmic loss at screen edges remains small (typically less than $1\%$) compared to diffraction loss.

These attractive features of the iris line were established by solving for the dominant eigenmodes supported by the structure. The found eigenmodes represent the stead-state solutions of the wave equation in a periodic geometry. As a boundary-value problem, it has been solved using the mode-matching technique assuming a Bloch-wave-like propagation \cite{NajiTHz1, Zotter} or using an equivalent impedance boundary condition (named Vainstein's boundary condition) that converts the problem into a smooth-pipe structure  \cite{DESY,Vainstein1,Vainstein2}. In such analyses, the iris-line structure is modeled approximately as an open resonator (with $r_{0}\rightarrow \infty$) and assumed to be infinitely long. 

In this paper, we perform a different type of field analysis, to solve for the fields in a line of finite length and to visualize any transient regime that may exist near the entrance of the line.  In other words, we solve the `forward' problem, rather than the settled stead-state problem.  Using the example of THz pump-probe experiments as a typical application \cite{DESY,NajiTHz1,Zhang}, where the THz pulse width is taken to be much smaller than the repetition rate, we can assume a single THz pulse passage through the structure.  The driving source is assumed to be of dipole polarization (e.g.~linear THz undulator \cite{NajiTHz1,Zhang}) and we exploit the fact that the structure is oversized (overmoded) to make simplifying assumptions in the field analysis.  Specifically, since the dimensions $a,b$ are much larger than the THz wavelength ($\lambda$) and the THz pulse width, a paraxially incident wave is assumed to have a slowly-varying envelope and a predominantly forward-scattered propagation along the axial ($z$) direction. The forward-scatter analysis presented uses TE-TM orthogonal decompositions to represent any arbitrary wave or geometrical discontinuity in the iris line. Given the periodicity of the iris line, we only have to work out the decompositions for one cell (period) of the line, modelling it as a forward-scatter matrix, to deduce the behaviour of the entire line as a cascade of such cells. 

The diffraction behavior trends obtained from this approach are shown to correlate well with the predictions made by the Vainstein boundary condition method and the perturbative mode-matching method, \cite{DESY, NajiTHz1}, but with an order-of-magnitude improvement in computational speed compared to the latter and with the ability to reveal transients along the line for arbitrary source (excitation) profiles. The approach is easily extendable to include tolerance sensitivity tests, whereby mechanical defects or aperiodic effects in specific cells can be inserted stochastically into the iris line. This is a natural result of formulating the problem in terms of scattering models (matrices).

The paper is organized as follows. Starting from first principles, Section~\ref{sec:formalism} builds the theoretical formulation of the problem and develops its approximate forward-scatter field solution. An orthogonal basis for modal decomposition inside the iris-line cells is developed, which is then used to represent the discontinuities of the geometry. Section~\ref{sec: implementation} implements the theoretical model numerically using forward-scatter transmission matrices and applies the method to several numerical examples, each with a different excitation source. The performance of the presented method is discussed and compared to the methods based on Vainstein's boundary condition and on perturbative mode matching. The paper is concluded in Section~\ref{sec:conclusions}, followed by two Appendices that contain several derivations for wave solutions and integrals needed in Section~\ref{sec:formalism}.

Throughout the paper we assume a time-harmonic dependence of the form $e^{-i\omega t}$, where $i=\sqrt{-1}$, $t$ is the time and $\omega$ is the angular frequency.


\section{\label{sec:formalism} Analytical Formulation}

\subsection{\label{sec:BasicFields} Orthogonal basis representation in each cell}

\begin{figure}
	\includegraphics[width=0.95\columnwidth]{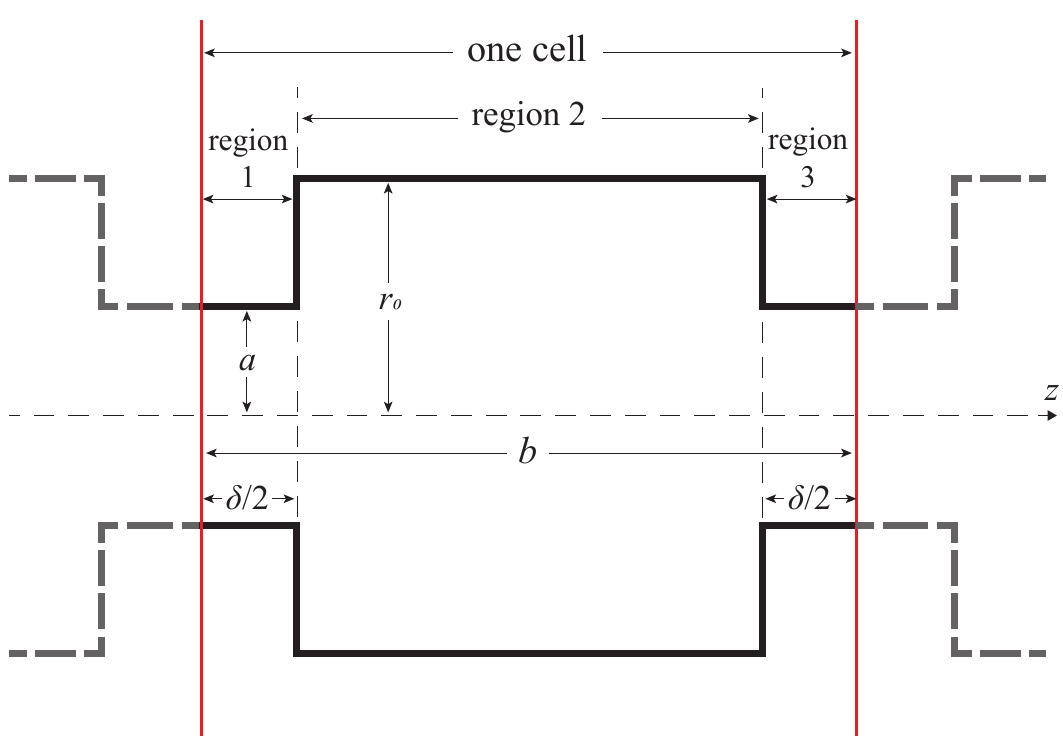}
	\caption{One cell of the iris line and regions of solution.}
	\label{fig:cell}
\end{figure}

Fig.~\ref{fig:cell} shows the geometry of each cell in the iris line. For convenience and conformity with the literature (e.g.~\cite{Stupakov2006}), we will use the following terminology: the short regions (regions 1 and 3) at the entrance and exit of each cell are referred to as `waveguide' sections, whereas the wider middle region (region 2) is referred to as the `cavity' section; the transition (discontinuity) from region 1 to region 2 is referred to as a `step-out' transition, whereas the the transition from region 2 to region 3 is referred to as a `step-in' transition. Given the axisymmetric nature of the geometry, we work chiefly in the cylindrical frame of coordinates, denoted by $(r,\phi,z)$. 

In this subsection we summarize the paraxial TE-TM mode expansions for all the components of a dipole field in a smooth circular pipe section, which is applicable to both the waveguide and cavity sections of our cell. Full derivations are provided in Appendix~\ref{Appx1}, starting from the wave equation and using the transverse field components $(E_{r},E_{\phi})$ as the main variables, for convenience. Note that although the scaling and derivation steps used may appear somewhat different from the traditional approach that relies on deriving the fields from their longitudinal components $(E_{z},H_{z})$ in canonical TE/TM waveguides (e.g.~\cite{Collin,pozar,jackson}), the final results can be shown to be fundamentally equivalent under the same paraxial propagation assumptions. Indeed, it is known from Whittaker's theorem that we can solve the  wave equation (i.e.~all 6 components of the electric and magnetic field) in terms of any two field components \cite{zangwill}.  

The following equations represent the final field expressions for the TE and TM modes in the waveguide section, with similar expressions found for the cavity section after replacing the radius $a$ by $r_{0}$. We expect a mode that is incident from the input waveguide (region 1) upon the step-out transition leading to the cavity (region 2) to be expanded as an infinite series of cavity modes. To signify this, the fields are written below as infinite sums in the cavity region. Note, however, that a parallel situation exists between a cavity mode incident upon the step-in transition, which is then decomposed into a series of waveguide modes in the output waveguide (region 3). In all the following equations, $Z_{0}$ denotes the free-space wave impedance, $\nu_{ij}$ denotes the $j$th zero of the Bessel function of the first kind and $i$th order ($J_{i}$) and $\nu'_{ij}$ denotes the $j$th zero of the derivative of the Bessel function of the first kind and $i$th order ($J'_{i}$).

\noindent TE $n$th mode in wavgeuide:
\begin{eqnarray}
   E_{ r}&=&\frac{A_{n}a}{\nu'_{1n} r} J_{1}(\nu'_{1n}\frac{ r}{a})\cos\phi \ e^{-iz\frac{\nu'^{2}_{1n}}{2ka^{2}}+ikz}\label{ArhoTE}\\
   E_{\phi}&=& -A_{n} J'_{1}(\nu'_{1n}\frac{ r}{a})\sin\phi \ e^{-iz\frac{\nu'^{2}_{1n}}{2ka^{2}}+ikz}\label{AphiTE}\\
   E_{z}&=&0 \label{AzTE}\\
   H_{ r}&=&\frac{-A_{n}}{Z_{0}}\left(\frac{\nu'^{2}_{1n}}{2k^{2}a^{2}}-1 \right)J'_{1}(\nu'_{1n}\frac{ r}{a})\sin\phi \ e^{-iz\frac{\nu'^{2}_{1n}}{2ka^{2}}+ikz}\label{HrTE}\\
   H_{\phi}&=&\frac{-A_{n}a}{Z_{0} r\nu'_{1n}}\left(\frac{\nu'^{2}_{1n}}{2k^{2}a^{2}}-1 \right)J_{1}(\nu'_{1n}\frac{ r}{a})\cos\phi \ e^{-iz\frac{\nu'^{2}_{1n}}{2ka^{2}}+ikz}\label{HphiTE}\\
   H_{z}&=&\frac{-iA_{n}\nu'_{1n}}{Z_{0}k a}J_{1}(\nu'_{1n}\frac{ r}{a})\sin\phi \ e^{-iz\frac{\nu'^{2}_{1n}}{2ka^{2}}+ikz}\label{HzTE}
\end{eqnarray}

\noindent TE modal expansion in cavity:
\begin{eqnarray}
   E_{ r}&=&\sum\limits_{n=1}^{\infty}\frac{A_{n}r_{0}}{\nu'_{1n} r} J_{1}(\nu'_{1n}\frac{ r}{r_{0}})\cos\phi \ e^{-iz\frac{\nu'^{2}_{1n}}{2kr_{0}^{2}}+ikz}\label{ArhoTEc}\\
   E_{\phi}&=&\sum\limits_{n=1}^{\infty} -A_{n} J'_{1}(\nu'_{1n}\frac{ r}{r_{0}})\sin\phi \ e^{-iz\frac{\nu'^{2}_{1n}}{2kr_{0}^{2}}+ikz}\label{AphiTEc}\\
   E_{z}&=&0 \label{AzTEc}\\
   H_{ r}&=&\sum\limits_{n=1}^{\infty}\frac{-A_{n}}{Z_{0}}\left(\frac{\nu'^{2}_{1n}}{2k^{2}r_{0}^{2}}-1 \right)J'_{1}(\nu'_{1n}\frac{ r}{r_{0}}) \nonumber\\
   && \times \sin\phi \ e^{-iz\frac{\nu'^{2}_{1n}}{2kr_{0}^{2}}+ikz}\label{HrTEc}\\
   H_{\phi}&=&\sum\limits_{n=1}^{\infty}\frac{-A_{n}r_{0}}{Z_{0} r\nu'_{1n}}\left(\frac{\nu'^{2}_{1n}}{2k^{2}r_{0}^{2}}-1 \right)J_{1}(\nu'_{1n}\frac{ r}{r_{0}}) \nonumber\\
   && \times \cos\phi \ e^{-iz\frac{\nu'^{2}_{1n}}{2kr_{0}^{2}}+ikz}\label{HphiTEc}\\
   H_{z}&=&\sum\limits_{n=1}^{\infty}\frac{-iA_{n}\nu'_{1n}}{Z_{0}k r_{0}}J_{1}(\nu'_{1n}\frac{ r}{r_{0}})\sin\phi \ e^{-iz\frac{\nu'^{2}_{1n}}{2kr_{0}^{2}}+ikz}\label{HzTEc}
\end{eqnarray}

\noindent TM $n$th mode in waveguide:
\begin{eqnarray}
       E_{ r}&=&-B_{n}J'_{1}\left( \nu_{1n}\frac{ r}{a}\right)\cos\phi \ e^{-iz\frac{\nu^{2}_{1n}}{2ka^{2}}+ikz}\label{ArhoTM}\\
       E_{\phi}&=& \frac{B_{n}a}{\nu_{1n} r}J_{1}\left( \nu_{1n}\frac{ r}{a}\right)\sin\phi \ e^{-iz\frac{\nu^{2}_{1n}}{2ka^{2}}+ikz}\label{AphiTM}\\
       E_{z}&=&\frac{+iB_{n}\nu_{1n}}{ak}J_{1}\left( \nu_{1n}\frac{ r}{a} \right)\cos\phi \ e^{-iz\frac{\nu^{2}_{1n}}{2ka^{2}}+ikz}\label{AzTM}\\
       H_{ r}&=&\frac{-B_{n}}{Z_{0} r}\left(\frac{\nu_{1n}}{2ak^{2}}+\frac{a}{\nu_{1n}} \right)J_{1}(\nu_{1n}\frac{ r}{a})\sin\phi \ e^{-iz\frac{\nu^{2}_{1n}}{2ka^{2}}+ikz}\label{HrTM}\\
       H_{\phi}&=&\frac{-B_{n}}{Z_{0}}\left(1+\frac{\nu^{2}_{1n}}{2a^{2}k^{2}}\right)J'_{1}(\nu_{1n}\frac{ r}{a})\cos\phi \ e^{-iz\frac{\nu^{2}_{1n}}{2ka^{2}}+ikz}\label{HphiTM}\\
       H_{z}&=&0\label{HzTM}
\end{eqnarray}

\noindent TM modal expansion in cavity:
\begin{eqnarray}
  E_{ r}&=&-\sum\limits_{n=1}^{\infty}B_{n}J'_{1}\left( \nu_{1n}\frac{ r}{r_{0}}\right)\cos\phi \ e^{-iz\frac{\nu^{2}_{1n}}{2kr_{0}^{2}}+ikz}\label{ArhoTMc}\\
   E_{\phi}&=& \sum\limits_{n=1}^{\infty}\frac{B_{n}r_{0}}{\nu_{1n} r}J_{1}\left( \nu_{1n}\frac{ r}{r_{0}}\right)\sin\phi \ e^{-iz\frac{\nu^{2}_{1n}}{2kr_{0}^{2}}+ikz}\label{AphiTMc}\\
   E_{z}&=&\sum\limits_{n=1}^{\infty}\frac{+iB_{n}\nu_{1n}}{r_{0}k}J_{1}\left( \nu_{1n}\frac{ r}{r_{0}} \right)\cos\phi \ e^{-iz\frac{\nu^{2}_{1n}}{2kr_{0}^{2}}+ikz}\label{AzTMc}\\
    H_{ r}&=&\sum\limits_{n=1}^{\infty}\frac{-B_{n}}{Z_{0} r}\left(\frac{\nu_{1n}}{2r_{0}k^{2}}+\frac{r_{0}}{\nu_{1n}} \right)J_{1}(\nu_{1n}\frac{ r}{r_{0}}) \nonumber\\
    && \times \sin\phi \ e^{-iz\frac{\nu^{2}_{1n}}{2kr_{0}^{2}}+ikz}\label{HrTMc}\\
    H_{\phi}&=&\sum\limits_{n=1}^{\infty}\frac{-B_{n}}{Z_{0}}\left(1+\frac{\nu^{2}_{1n}}{2r_{0}^{2}k^{2}}\right)J'_{1}(\nu_{1n}\frac{ r}{r_{0}}) \nonumber\\
    && \times \cos\phi \ e^{-iz\frac{\nu^{2}_{1n}}{2kr_{0}^{2}}+ikz}\label{HphiTMc}\\
    H_{z}&=&0\label{HzTMc}
\end{eqnarray}

\subsection{\label{sec:CellScatteringModel} Derivation of the paraxial forward-scatter model for each cell of the iris line}

Using the orthogonal mode expansions developed in Section~\ref{sec:BasicFields}, we now derive the modal expansions that represent the step-out and step-in transitions within each cell, both for the case of an incident TE and an incident TM mode. From the results of such expansions we then build the paraxial forward-scatter model for each cell. 

\begin{figure}
	\includegraphics[width=0.9\columnwidth]{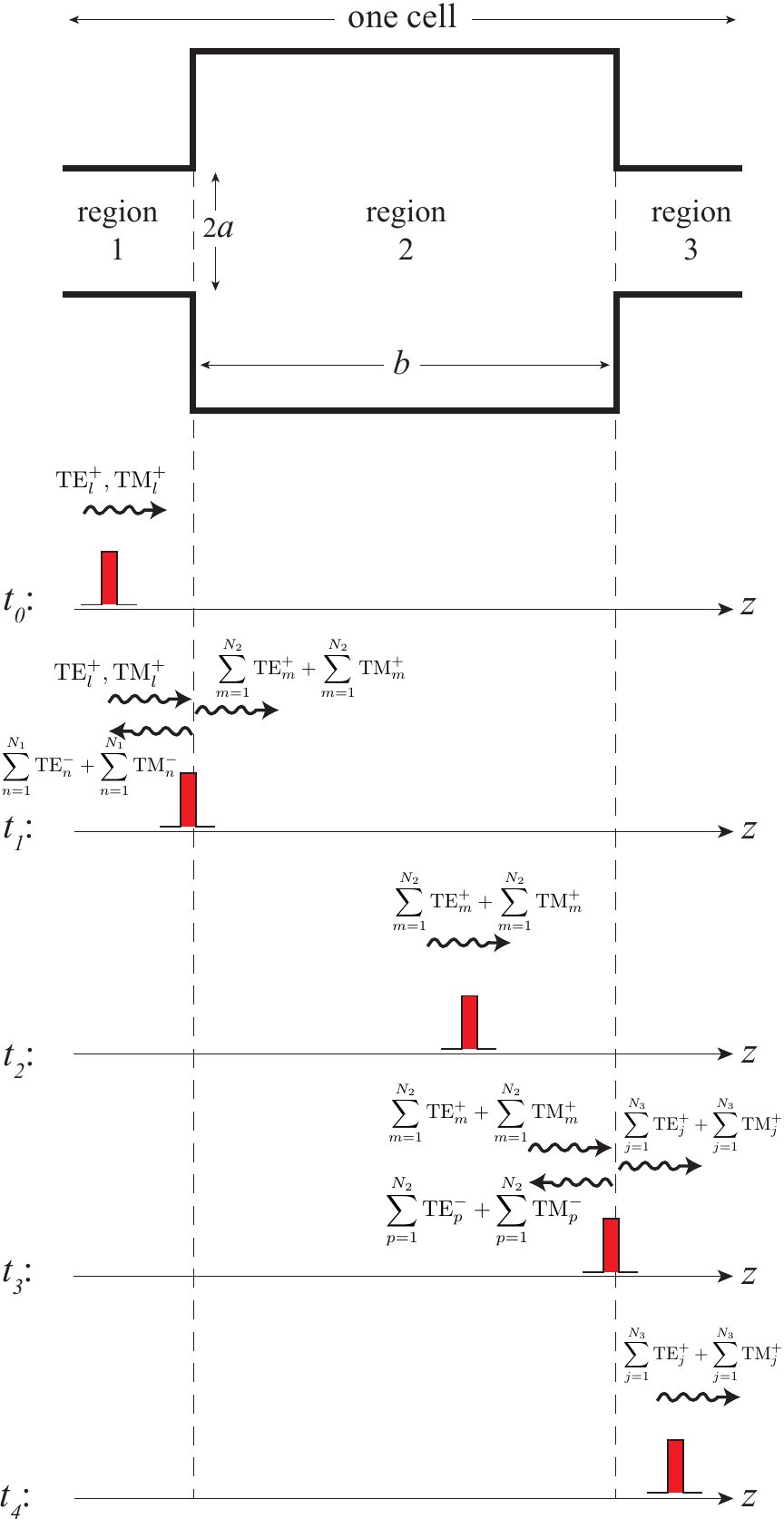}
	\caption{A qualitative sketch of a short THz pulse travelling through the oversized cell, with indicative forward and backward mode expansions at each boundary plane within the cell. For clarity, the pulse (in red) is shown at different moments in time.  At $t_{0}$ a forward-traveling wave is incident onto the cell. At $t_{1}$ the wave reaches the first discontinuity and we generally have: a forward mode and a backward scattered-mode expansion in region 1, along with a forward scattered-mode expansion in region 2 (due to the shortness of the pulse, no backward reflections exist yet in region 2). At $t_{2}$ the forward wave continues through region 2, and at $t_{3}$ the second discontinuity is met: again we generally have forward and backward scattered-mode expansions in region 2 and only a forward scattered-mode expansion in region 3. At $t_{4}$ forward traveling waves propagate in region 3 towards the cell's exit, and the cycle repeats through the next cells. A key assumption made in the forward-scatter high-frequency analysis under consideration (see the main text) is that we can expand the wave in forward modes only, ignoring the backward reflections as a first approximation, which turns out to be a good approximation at the high-frequency limit.}
	\label{fig:strategy}
\end{figure}

We shall simplify the scattering model of the cell by exploiting the fact that the iris-line cell dimensions ($a,b$) are much larger than the THz wavelength ($\lambda$) and the paraxial pulse width. Fig.~\ref{fig:strategy} depicts how one may generally decompose the electromagnetic wave travelling in each section of the cell using forward- and backward-traveling mode expansions. In this analysis we ignore the local backward reflections or standing waves created by the oversized step-in/out transitions and expand the fields in terms of forward modes only (symbolized in the Fig.~\ref{fig:strategy} by positive superscripts) when matching boundary conditions at the transitions. One can envisage a scenario in which multiple reflections between the screens in a closed-chamber setup could lead to the formation of echo-like pulses that trail the primary pulse after some delay. However, such a delay will be at least the time required to travel the round-trip distance $2b$, which is assumed to be much longer than the typical interval between a THz pump pulse and its corresponding x-ray probe pulse (in typical pump-probe experiments). We therefore confine our attention in this analysis to the single-pulse passage scenario, with no secondary pulses or echoes.

\subsubsection{\label{TM_step-out} Incident TM on a step-out discontinuity} 

Consider a TM mode (call it the $l$-th mode) incident from the left upon a the step-out transition, with $z=0$ set for convenience as the location of the transition. This can represent a mode lauched in the input waveguide (region 1) and travelling towards the cavity (region 2).  From (\ref{ArhoTM}) and (\ref{AphiTM}) we have the radial and azimuthal components of the electric field in the region $z<0$ given by 
  \begin{eqnarray}\label{eq:20}
  E_{r}& =&-B_{l}J'_{1}\left( \nu_{1l}\frac{r}{a}\right)   \cos\phi \, e^{-iz\frac{\nu^{2}_{1l}}{2ka^{2}}},  \\
  E_{\phi}&=&   \frac{B_{l}a}{\nu_{1l}r}   J_{1}\left( \nu_{1l}\frac{r}{a}\right)   \sin\phi \,   e^{-iz\frac{\nu^{2}_{1l}}{2ka^{2}}},
  \end{eqnarray}
which will be expanded into a sum of TE and TM modes in region $z>0$, when matched at the transition's plane ($z=0$), as implied by (\ref{ArhoTEc})--(\ref{HzTEc}) and (\ref{ArhoTMc})--(\ref{HzTMc}). We thus write the continuity of the radial fields as 
  \begin{equation}\label{eq:21}
  -B_{l}J'_{1}\left( \nu_{1l}\frac{r}{a}\right)
  =
  \sum\limits_{n=1}^{\infty}
  \frac{A_{n}r_{0}}{\nu'_{1n}r} J_{1}(\nu'_{1n}\frac{r}{r_{0}}) 
  {-}
  \sum\limits_{n=1}^{\infty}
  B_{n}J'_{1}\left( \nu_{1n}\frac{r}{r_{0}}\right),
  \end{equation}
and the continuity of the azimuthal fields as 
  \begin{equation}\label{eq:22}
  \frac{B_{l}a}{\nu_{1l}r}
  J_{1}\left( \nu_{1l}\frac{r}{a}\right)
  =  {-}   \sum\limits_{n=1}^{\infty} 
  A_{n} J'_{1}(\nu'_{1n}\frac{r}{r_{0}})
  +
  \sum\limits_{n=1}^{\infty}
  \frac{B_{n}r_{0}}{\nu_{1n}r}J_{1}\left( \nu_{1n}\frac{r}{r_{0}}\right).
  \end{equation}

We now proceed to find coefficients $A_n$ and $B_n$ analytically, which would characterize the step-out transition's effect on incident TM waves. Multiplying \eqref{eq:21} by $J'_{1}(\nu_{1m}{r}/{r_{0}})$ and integrating across with $\int_0^{r_{0}} r\,dr$, which reduces to $\int_0^a r\,dr$ on the lhs (since the radial field is zero for $a<r<r_{0}$), gives  
  \begin{eqnarray}\label{eq:23}
  &&-B_{l}
  \int_0^a r\,dr
  J'_{1}\left( \nu_{1l}\frac{r}{a}\right)
  J'_{1}(\nu_{1m}\frac{r}{r_{0}}) \nonumber\\
  &=&  -
  \sum\limits_{n=1}^{\infty}
  \frac{A_{n}r_{0}}{\nu_{1m}} 
  \int_0^{r_{0}} dr
  J'_{1}\left(\nu'_{1n}\frac{r}{r_{0}}\right) 
  J_{1}\left(\nu_{1m}\frac{r}{r_{0}}\right)\nonumber\\
  &&-
  \sum\limits_{n=1}^{\infty}
  B_{n}
  \int_0^{r_{0}} r\,dr J'_{1}\left( \nu_{1n}\frac{r}{r_{0}}\right)
  J'_{1}\left(\nu_{1m}\frac{r}{r_{0}}\right),
  \end{eqnarray}
where we have used integration by parts to condition the first sum on the rhs. Similarly, multiplying \eqref{eq:22} by $({r_{0}}/{\nu_{1m} r})J_{1}\left( \nu_{1m}\frac{r}{r_{0}}\right)$ and integrating across with $\int_0^{r_{0}} r\,dr$, which reduces to $\int_0^a r\,dr$ on the lhs (since the azimuthal field is zero for $a<r<r_{0}$), gives
  \begin{eqnarray}\label{eq:24}
  &&\frac{B_{l}a}{\nu_{1l}}
  \frac{r_{0}}{\nu_{1m}}
  \int_0^a \frac{dr}{r}
  J_{1}\left( \nu_{1l}\frac{r}{a}\right)
  J_{1}\left( \nu_{1m}\frac{r}{r_{0}}\right)\nonumber\\
  &=&
  -
  \sum\limits_{n=1}^{\infty} 
  \frac{A_{n}r_{0}}{\nu_{1m}} \int_0^{r_{0}} dr
  J'_{1}\left(\nu'_{1n}\frac{r}{r_{0}}\right)
  J_{1}\left( \nu_{1m}\frac{r}{r_{0}}\right)
  \nonumber\\&&
  +
  \sum\limits_{n=1}^{\infty}
  \frac{B_{n}r_{0}}{\nu_{1n}}
  \frac{r_{0}}{\nu_{1m}}
  \int_0^{r_{0}} \frac{dr}{r}
  J_{1}\left( \nu_{1n}\frac{r}{r_{0}}\right)
  J_{1}\left( \nu_{1m}\frac{r}{r_{0}}\right). 
  \end{eqnarray}
  
We can now eliminate the coefficients $A_n$ and find coefficients $B_n$ by subtracting \eqref{eq:24} from \eqref{eq:23}, to yield
  \begin{equation}
  -B_{l}(I_{3})= -\sum\limits_{n=1}^{\infty} B_{n}(I_{1}), \label{eq:25}
  \end{equation}
  where $I_{1}$ and $I_{3}$ denote the integrals
  \begin{eqnarray}
 I_{1} &=&
  \int_0^{r_{0}} r\,dr J'_{1}\left( \nu_{1n}\frac{r}{r_{0}}\right)
  J'_{1}\left(\nu_{1m}\frac{r}{r_{0}}\right) \nonumber\\
  && +
  \frac{r_{0}^2}{\nu_{1n}\nu_{1m}}
  \int_0^{r_{0}} \frac{dr}{r}
  J_{1}\left( \nu_{1n}\frac{r}{r_{0}}\right)
  J_{1}\left( \nu_{1m}\frac{r}{r_{0}}\right),\label{I1_first}\\
 I_{3} &=& \int_0^a r\,dr
  J'_{1}\left( \nu_{1l}\frac{r}{a}\right)
  J'_{1}(\nu_{1m}\frac{r}{r_{0}}) \nonumber\\
  &&+
  \int_0^a dr
  \frac{a}{\nu_{1l}}
  J_{1}\left( \nu_{1l}\frac{r}{a}\right)
  \frac{r_{0}}{\nu_{1m}r}J_{1}\left( \nu_{1m}\frac{r}{r_{0}}\right).  \label{I3_first} 
  \end{eqnarray}

Analytical calculation of integrals $I_{1}$ and $I_{3}$ can be found in Appendix~\ref{Appx2}. Substituting from \eqref{I1_result} and \eqref{I3_result} into \eqref{eq:25} gives the final result as
 \begin{equation}\label{eq:242}
  B_m =  \frac{-2B_{l}a\nu_{1m}J_{0}(\nu_{1l})J_{1}\left( \nu_{1m}\frac{a}{r_{0}}\right)}{r_{0}^3 J'_{1}\left( \nu_{1m}\right)^2\left(\frac{\nu_{1l}^2}{a^2}-\frac{\nu_{1m}^2}{r_{0}^2}\right)} \ \ \text{(TM on step-out)}.
  \end{equation}

In a similar manner, we find coefficients $A_n$ by multiplying \eqref{eq:21} by $J_{1}(\nu'_{1m}{r}/{r_{0}})(r_{0}/\nu'_{1m}{r})$ and integrating across with $\int_0^{r_{0}} r\,dr$, which reduces to $\int_0^a r\,dr$ on the lhs, to give   
  \begin{eqnarray} \label{eq:28}
  &-&B_{l}
  \int_0^a r\,dr
  J'_{1}\left( \nu_{1l}\frac{r}{a}\right)
  J_{1}\left(\nu'_{1m}\frac{r}{r_{0}}\right)
  \frac{r_{0}}{\nu'_{1m}{r}}
  \nonumber\\
   &=&
  \sum\limits_{n=1}^{\infty}
  \frac{A_{n}r_{0}}{\nu'_{1n}} 
  \int_0^{r_{0}} dr
  J_{1}\left(\nu'_{1n}\frac{r}{r_{0}}\right) 
  J_{1}\left(\nu'_{1m}\frac{r}{r_{0}}\right)
  \frac{r_{0}}{\nu'_{1m}{r}} \nonumber\\
  &&+
  \sum\limits_{n=1}^{\infty}
  B_{n}
  \frac{r_{0}}{\nu_{1n}}
  \int_0^{r_{0}} 
  dr J_{1}\left( \nu_{1n}\frac{r}{r_{0}}\right)
  J'_{1}\left(\nu'_{1m}\frac{r}{r_{0}}\right),
  \end{eqnarray}
where we have used integration by parts to condition the second sum on the rhs. Multiplying \eqref{eq:22} by $J'_{1}\left( \nu'_{1m}\frac{r}{r_{0}}\right)$ and integrating across with $\int_0^{r_{0}} r\,dr$, which reduces to $\int_0^a r\,dr$ on the lhs, similarly gives
  \begin{eqnarray}\label{eq:29}
  &&  \frac{B_{l}a}{\nu_{1l}}
  \int_0^a dr
  J_{1}\left( \nu_{1l}\frac{r}{a}\right)
  J'_{1}\left( \nu'_{1m}\frac{r}{r_{0}}\right)\nonumber\\
  &=& -
  \sum\limits_{n=1}^{\infty} 
  {A_{n}} 
  \int_0^{r_{0}} r dr
  J'_{1}\left(\nu'_{1n}\frac{r}{r_{0}}\right)
  J'_{1}\left( \nu'_{1m}\frac{r}{r_{0}}\right)
  \nonumber\\
  && +
  \sum\limits_{n=1}^{\infty}
  \frac{B_{n}r_{0}}{\nu_{1n}}
  \int_0^{r_{0}} dr
  J_{1}\left( \nu_{1n}\frac{r}{r_{0}}\right)
  J'_{1}\left( \nu'_{1m}\frac{r}{r_{0}}\right).
  \end{eqnarray}
  
To eliminate $B_{n}$ and find $A_{n}$ we subtract \eqref{eq:29} from \eqref{eq:28}, to yield
  \begin{equation}\label{eq:30}
  -B_{l} (I_{4})=\sum\limits_{n=1}^{\infty} A_{n} (I_{2}),
  \end{equation}
where $I_{2}$ and $I_{4}$ denote the integrals
  \begin{eqnarray}
  I_{2}&=&\frac{r_{0}}{\nu'_{1n}} 
  \int_0^{r_{0}} dr
  J_{1}\left(\nu'_{1n}\frac{r}{r_{0}}\right) 
  J_{1}\left(\nu'_{1m}\frac{r}{r_{0}}\right)
  \frac{r_{0}}{\nu'_{1m}{r}}\nonumber\\
  && +   \int_0^{r_{0}} r dr
  J'_{1}\left(\nu'_{1n}\frac{r}{r_{0}}\right)
  J'_{1}\left( \nu'_{1m}\frac{r}{r_{0}}\right)\label{I2_first}\\
  I_{4}&=&\frac{r_{0}}{\nu'_{1m}}
  \int_0^a  dr   J'_{1}\left( \nu_{1l}\frac{r}{a}\right)
  J_{1}\left(\nu'_{1m}\frac{r}{r_{0}}\right) \nonumber\\
  &&+
  \frac{a}{\nu_{1l}}
  \int_0^a dr
  J_{1}\left( \nu_{1l}\frac{r}{a}\right)
  J'_{1}\left( \nu'_{1m}\frac{r}{r_{0}}\right) \label{I4_first}
  \end{eqnarray}

We notice that the integrand in $I_{4}$ is actually the perfect differential
\begin{equation}
  \frac{r_{0}}{\nu'_{1m}}
  \frac{a}{\nu_{1l}}
  \frac{d}{dr}
  \left[
  J_{1}\left( \nu_{1l}\frac{r}{a}\right)
  J_{1}\left(\nu'_{1m}\frac{r}{r_{0}}\right)
  \right],
  \end{equation}
from which the integral $I_{4}$ is immediately found to be zero. Integral $I_{2}$ is calculated in Appendix~\ref{Appx2} and its result is substituted from \eqref{I2_result} into \eqref{eq:30}, to give the final result 
\begin{equation}\label{eq:AmTMStepOut}
    A_{m}=0\ \ \ \text{(TM on step-out)}.
\end{equation}

\begin{figure}
	\includegraphics[width=0.8\columnwidth]{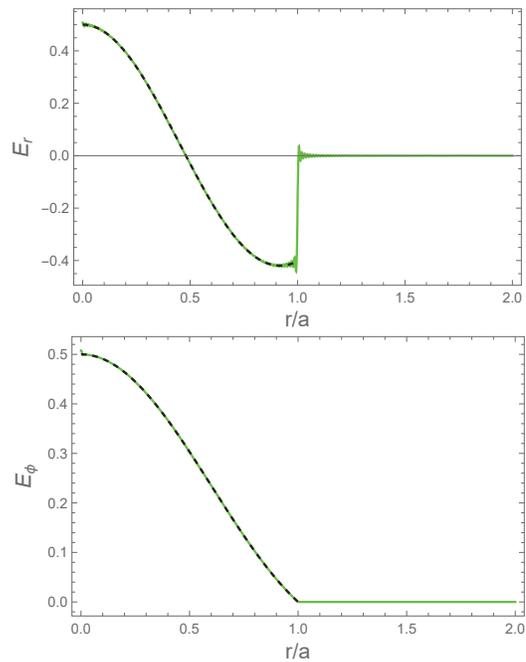}
	\caption{Comparison of the right-hand sides (green color line) with the left-hand sides (dashed black line) of equation \eqref{eq:21} for $E_{r}$ and equation \eqref{eq:22} for $E_{\phi}$, with the upper limit of the sums truncated at 300 terms and the $(A_{n},B_{n})$ values substituted from the results in \eqref{eq:AmTMStepOut} and \eqref{eq:242}. In this plot we took $r_{0}/a=2$ as an example.}
	\label{fig:TMonStepOutBoth}
\end{figure}

In Fig.~\ref{fig:TMonStepOutBoth} we compare the right-hand sides of \eqref{eq:21} and \eqref{eq:22} with the corresponding left-hand sides, using the results in \eqref{eq:242} and \eqref{eq:AmTMStepOut} and with the sums truncated to $n=300$ terms. Good agreement is observed between the original mode and its decomposition. A local oscillatory effect is noticed near the edge of the pipe ($r\rightarrow a$), as expected from the Fresnel diffraction. Interestingly, it can be shown that this local oscillation near the edge, where the parameter $r/a-1\equiv \tau$ is small in magnitude, is asymptotically well approximated (after removing the average value) by the form ${\sim}\text{Erf}\left( \tau\sqrt{ka^{2}/2iz}-\tau \right)$, where $\text{Erf($\cdot$)}$ denotes the Error Function. We shall leave the details of this asymptotic analysis for a future publication.

\subsubsection{\label{TM_step-in} Incident TM on a step-in discontinuity} 

We now repeat the analysis done in Subsection~\ref{TM_step-out}, but for the case of a wave incident from the cavity (region 2) into the narrower output waveguide (region 3). Specifically, we now have the following radial and azimuthal matching conditions
\begin{equation}
    B_{l}J'_{1}\left( \nu_{1l}\frac{r}{r_{0}}\right)=
  \sum\limits_{n=1}^{\infty}
  -\frac{A_{n}a}{\nu'_{1n}r} J_{1}(\nu'_{1n}\frac{r}{a}) 
  + \sum\limits_{n=1}^{\infty}
  B_{n}J'_{1}\left( \nu_{1n}\frac{r}{a}\right)\label{eq:21New}
  \end{equation}
  \begin{equation}
  \frac{B_{l}r_{0}}{\nu_{1l}r}
  J_{1}\left( \nu_{1l}\frac{r}{r_{0}}\right) =
  -  \sum\limits_{n=1}^{\infty} 
  A_{n} J'_{1}(\nu'_{1n}\frac{r}{a})
  +  \sum\limits_{n=1}^{\infty}
  \frac{B_{n}a}{\nu_{1n}r}J_{1}\left( \nu_{1n}\frac{r}{a}\right)\label{eq:22New}
\end{equation}
Following similar steps to those taken above for \eqref{eq:30} (but multiplying with Bessel functions of arguments that contain $r/a$, instead of $r/r_{0}$, and integrating across with $\int^{a}_{0} r dr $ since we now are expanding in the waveguide region where there is no field beyond $r=a$), we solve for $A_{n}$ and obtain
  \begin{equation}\label{eq:32New}
      A_{m} (\hat{I}_{2})=-B_{l} (\hat{I}_{4}),
  \end{equation}
  where $\hat{I}_{2}$ and $\hat{I}_{4}$ are integrals with forms similar to $I_{2}$ and $I_{4}$ in \eqref{I2_first} and \eqref{I4_first}. Specifically, the integral $\hat{I}_{2}$ can be obtained from the integral $I_{2}$ in \eqref{I2_first} by interchanging the radii $r_{0} \leftrightarrow a$ in all terms and integration limits. As was shown in the calculation of \eqref{I2_result}, we can easily deduce that 
  \begin{equation}
      \hat{I}_{2}=\delta_{mn}\frac{a^{2}}{2}(1-1/\nu_{1m}^{2})J^{2}_{1}(\nu'_{1m}), \label{newQmEq}
  \end{equation}
where $\delta_{mn}$ is Kronecker's delta function.

On the other hand, integral $\hat{I}_{4}$ can be obtained from integral $I_{4}$ in \eqref{I4_first} by exchanging the radii $r_{0} \leftrightarrow a$ in all terms, while keeping the integration limits from 0 to $a$; namely,
\begin{eqnarray}
\hat{I}_{4}&=&\frac{a}{\nu'_{1m}}
  \int_0^a   dr
  J'_{1}\left( \nu_{1l}\frac{r}{r_{0}}\right)
  J_{1}\left(\nu'_{1m}\frac{r}{a}\right) \nonumber\\
  &&+   \frac{r_{0}}{\nu_{1l}}
  \int_0^a dr
  J_{1}\left( \nu_{1l}\frac{r}{r_{0}}\right)
  J'_{1}\left( \nu'_{1m}\frac{r}{a}\right)\nonumber\\
   &=&\frac{r_{0}a J_{1}(\nu_{1l}\frac{a}{r_{0}})J_{0}(\nu'_{1m})}{\nu_{1l}},\label{resultforAm}
\end{eqnarray}
where in the last step we used the fact that $J_{1}(\nu'_{1m})=\nu'_{1m}J_{0}(\nu'_{1m})$, since $J'_{1}(\nu'_{1m})=0$ by definition.

Substituting the results from \eqref{newQmEq} and \eqref{resultforAm} back into \eqref{eq:32New} and simplifying give the value for $A_{n}$ as
\begin{equation}
    A_{m}=\frac{-2B_{l} r_{0}J_{1}(\nu_{1l}a/r_{0})}{a \nu_{1l}\left(\nu'_{1m}{-}\frac{1}{\nu'_{1m}}\right)J_{1}(\nu'_{1m})} \ \ \text{(TM on step-in)}.\label{NewAm_eq}
\end{equation}
\begin{figure}
	\includegraphics[width=0.8\columnwidth]{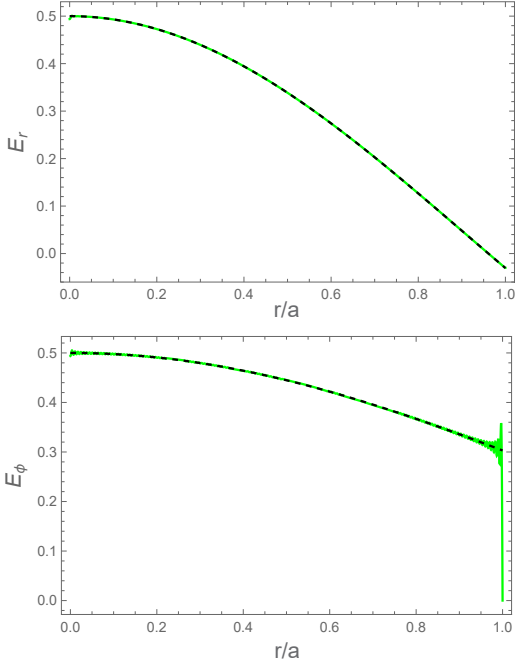}
	\caption{Comparison of the right-hand sides (green color line) with the left-hand sides (dashed black line) of equation \eqref{eq:21New} for $E_{r}$ and equation \eqref{eq:22New} for $E_{\phi}$, with the upper limit of the sums truncated at 300 terms and the $(A_{n},B_{n})$ values substituted from the results in \eqref{NewAm_eq} and \eqref{NewBm_eq}. In this plot we took $r_{0}/a=2$ as an example.}
	\label{fig:TMonStepInBoth}
\end{figure}
It is worth noting that $A_{m}\neq 0$ here, unlike the case for the step-out discontinuity. This indicates that TE modes (not only the TM modes) will be expanded in the waveguide following a TM incidence on the step-in discontinuity.

Coefficients $B_{n}$ are obtained in a similar way, which parallels that used to reach \eqref{eq:25} from Subsection~\ref{TM_step-out}, leading to
  \begin{equation}\label{eq:32New__}
      B_{m} (\hat{I}_{1})=B_{l} (\hat{I}_{3}),
  \end{equation}
  where $\hat{I}_{1}$ and $\hat{I}_{3}$ are integrals with forms similar to $I_{1}$ and $I_{3}$ in \eqref{I1_first} and \eqref{I3_first}. Specifically, the integral $\hat{I}_{1}$ can be obtained from the integral $I_{1}$ in \eqref{I1_first} by interchanging the radii $r_{0} \leftrightarrow a$ in all terms and integration limits. As was shown in the calculation of \eqref{I1_result}, we can easily deduce that 
\begin{equation}
      \hat{I}_{1}=\delta_{mn}\frac{a^{2}}{2}J'^{2}_{1}(\nu_{1m}). \label{NewRm_eq}
\end{equation}

Integral $\hat{I}_{3}$ can be obtained from integral $I_{3}$ in \eqref{I3_first} by exchanging the radii $r_{0} \leftrightarrow a$ in all terms, while keeping the integration limits from 0 to $a$; namely,
\begin{eqnarray}
\hat{I}_{3}&=&\int_0^a dr\left[r
  J'_{1}\left( \nu_{1l}\frac{r}{r_{0}}\right)
  J'_{1}(\nu_{1m}\frac{r}{a}) \right. \nonumber\\
  && \left.  +
  r\frac{r_{0}}{\nu_{1l}r}
  J_{1}\left( \nu_{1l}\frac{r}{r_{0}}\right)
  \frac{a}{\nu_{1m}r} J_{1}\left( \nu_{1m}\frac{r}{a}\right)
  \right]\nonumber\\
  &=&\frac{a \nu_{1l} J_{0}(\nu_{1m})J_{1}(\frac{\nu_{1l}a}{r_{0}})}{r_{0}(\nu^{2}_{1l}/r_{0}^{2}-\nu^{2}_{1m}/a^{2})}, \label{newI3Eq}
    \end{eqnarray}
noting that this result has $\nu_{1m} \leftrightarrow \nu_{1l}$ interchanged, compared to the result of $I_{3}$ in \eqref{I3_result}. The value of $B_{n}$ is now found by substituting \eqref{newI3Eq}
 and \eqref{NewRm_eq} into \eqref{eq:32New__} and simplifying, to yield
 \begin{equation}
     B_{m}=\frac{-2B_{l}a\nu_{1l}J_{0}(\nu_{1m})J_{1}(\nu_{1l}\frac{a}{r_{0}})}{r_{0}(\nu^{2}_{1m}-\nu^{2}_{1l}\frac{a^{2}}{r_{0}^{2}})J'^{2}_{1}(\nu_{1m})} \ \ \text{(TM on step-in)}. \label{NewBm_eq}
 \end{equation}

In Fig.~\ref{fig:TMonStepInBoth} we compare the right-hand sides of \eqref{eq:21New} and \eqref{eq:22New} with the corresponding left-hand sides, using the results in \eqref{NewAm_eq} and \eqref{NewBm_eq} and with the sums truncated to $n=300$ terms. Good agreement is observed between the original mode and its decomposition.

\subsubsection{\label{TE_step-out} Incident TE on a step-out discontinuity} 

Let us now consider a step-out discontinuity with an incident TE mode (incidence from waveguide towards cavity). At the discontinuity location, again taken as $z=0$ for convenience, the matched radial electric field gives
\begin{equation}
    \frac{A_{l}a}{\nu'_{1l} r}J_{1}(\nu'_{1l}\frac{r}{a})=\sum\limits_{n}\frac{A_{n}r_{0}}{\nu'_{1n}r}J_{1}(\nu'_{1n}\frac{r}{r_{0}})-B_{n}J'_{1}(\nu_{1n}\frac{r}{r_{0}}),\label{TEmatch_Er}
\end{equation}
while the matched azimuthal electric field gives
\begin{equation}
    A_{l}J'_{1}(\nu'_{1l}\frac{r}{a})=\sum\limits_{n}A_{n}J'_{1}(\nu'_{1n}\frac{r}{r_{0}})-\frac{B_{n}r_{0}}{\nu_{1n}r}J_{1}(\nu_{1n}\frac{r}{r_{0}}).\label{TEmatch_Ephi}
\end{equation}

To find coefficients $B_{n}$ we multiply \eqref{TEmatch_Ephi} by $J_{1}(\nu_{1m}r/r_{0})$ and integrate across with $\int^{r_{0}}_{0} r dr$, which is clipped to $\int^{a}_{0} r dr$ on the lhs for the waveguide fields, to yield
\begin{eqnarray}
    && A_{l}\int^{a}_{0}J'_{1}(\nu'_{1l}\frac{r}{a})J_{1}(\nu_{1m}\frac{r}{r_{0}}) dr\nonumber\\
    &&=\sum\limits_{n}A_{n}\int^{r_{0}}_{0}J'_{1}(\nu'_{1n}\frac{r}{r_{0}})J_{1}(\nu_{1m}\frac{r}{r_{0}})dr\nonumber\\
    &&-\int^{r_{0}}_{0}\frac{B_{n}r_{0}}{\nu_{1n}r}J_{1}(\nu_{1n}\frac{r}{r_{0}})J_{1}(\nu_{1m}\frac{r}{r_{0}})dr.\label{TEmatch_EphiNew1}
\end{eqnarray}

Similarly, we multiply \eqref{TEmatch_Er} by $\frac{-\nu_{1m}r}{r_{0}}J'_{1}(\nu_{1m}r/r_{0})$, integrate, then use integration by parts on the first term on the rhs to obtain
\begin{eqnarray}
    &&\frac{-A_{l}a\nu_{1m}}{\nu'_{1l}r_{0}}\int^{a}_{0}J_{1}(\nu'_{1l}\frac{r}{a})J'_{1}(\nu_{1m}\frac{r}{r_{0}})dr\nonumber\\
    &&=\sum\limits_{n} A_{n}\int^{r_{0}}_{0}J_{1}(\nu_{1m}\frac{r}{r_{0}})J'_{1}(\nu'_{1n}\frac{r}{r_{0}}) dr\nonumber\\
    &&+B_{n}\frac{\nu_{1m}}{r_{0}}\int^{r_{0}}_{0}rJ'_{1}(\nu_{1n}\frac{r}{r_{0}})J'_{1}(\nu_{1m}\frac{r}{r_{0}})dr.\label{TEmatch_ErNew2}
\end{eqnarray}

Subtracting \eqref{TEmatch_ErNew2} from \eqref{TEmatch_EphiNew1} now allows us to find $A_{n}$ as
\begin{equation}\label{TEStep-out_Bn_balance}
    A_{l} (L_{1})=\sum\limits_{n}-B_{n} (L_{2})
\end{equation}
where $L_{1}$ and $L_{2}$ are given by the following integrals
\begin{eqnarray}
   L_{1}&=&\int^{a}_{0} J'_{1}(\nu'_{1l} \frac{r}{a})J_{1}(\nu_{1m}\frac{r}{r_{0}})dr\nonumber\\
   &&+\frac{a\nu_{1m}}{r_{0} \nu'_{1l}}\int^{a}_{0}J_{1}(\nu'_{1l}\frac{r}{a})J'_{1}(\nu_{1m}\frac{r}{r_{0}})\label{L1_first} \\
   &=&aJ_{0}(\nu'_{1l})J_{1}(\nu_{1m}\frac{a}{r_{0}}), \label{L1_found}
    \end{eqnarray}
\begin{eqnarray}
   L_{2}&=&\int^{r_{0}}_{0} \frac{r_{0}}{\nu_{1n}r}J_{1}(\nu_{1n}\frac{r}{r_{0}})J_{1}(\nu_{1m}\frac{r}{r_{0}})dr \nonumber\\
    &&+\frac{\nu_{1m}}{r_{0}}\int^{r_{0}}_{0}rJ'_{1}(\nu_{1n}\frac{r}{r_{0}})J'_{1}(\nu_{1m}\frac{r}{r_{0}})dr \label{L2_first}\\
    &\equiv&\frac{\nu_{1m}}{r_{0}}I_{1}=\delta_{mn}\frac{r_{0}\nu_{1m}}{2}J^{2}_{0}(\nu_{1m}), \label{L2_found}
\end{eqnarray}
where, in \eqref{L1_found}, we used the fact that $J_{0}(\nu'_{1l})-J_{1}(\nu'_{1l})/\nu'_{1l}=J'_{1}(\nu'_{1l})=0$ by definition and hence $J_{1}(\nu'_{1l})/\nu'_{1l}=J_{0}(\nu'_{1l})$, and we substituted for $I_{1}$ from the former result in \eqref{I1_result}. 
\begin{figure}
	\includegraphics[width=0.8\columnwidth]{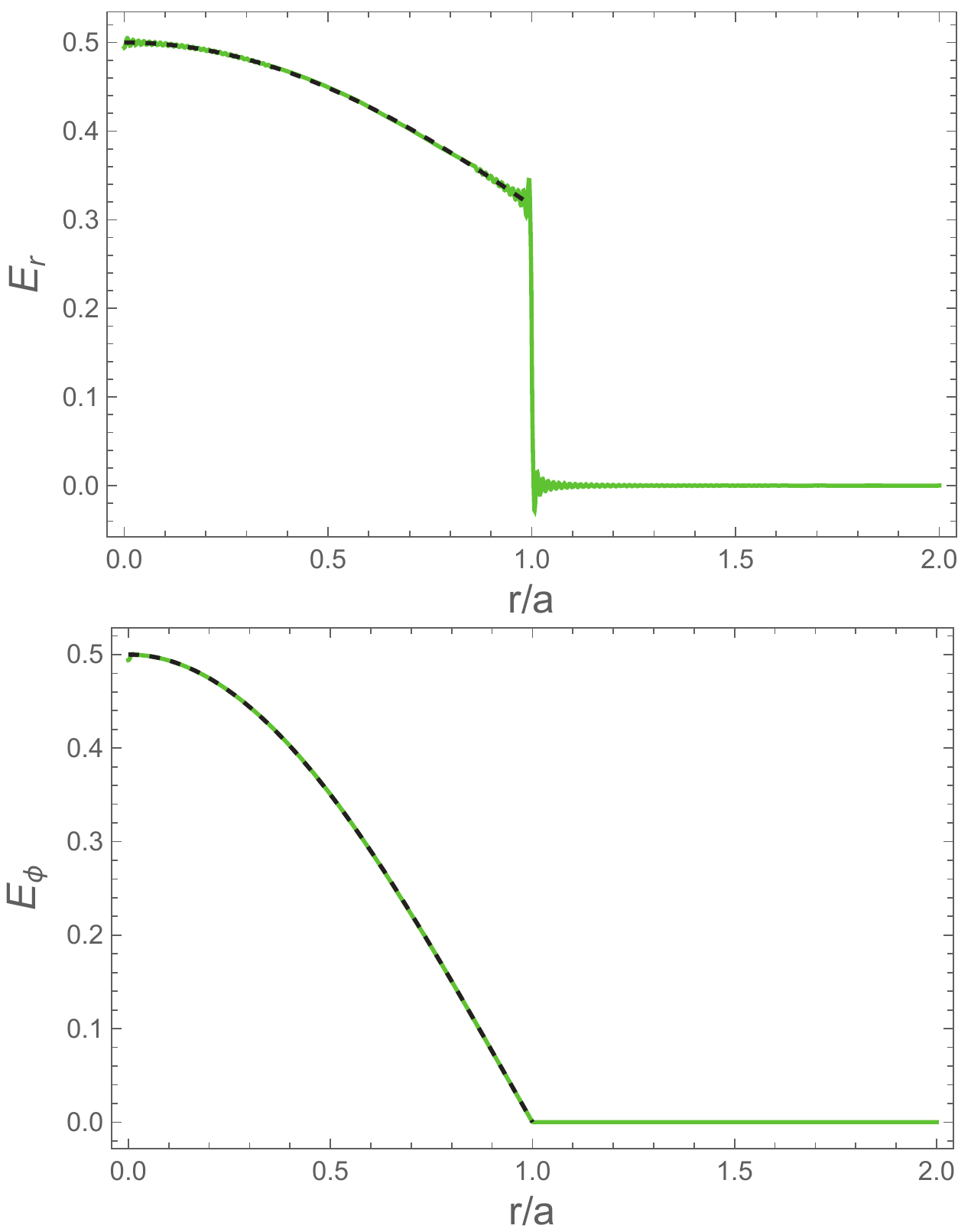}
	\caption{Comparison of the right-hand sides (green color line) with the left-hand sides (dashed black line) of equation \eqref{TEmatch_Er} for $E_{r}$ and equation \eqref{TEmatch_Ephi} for $E_{\phi}$, with the upper limit of the sums truncated at 300 terms and the $(A_{n},B_{n})$ values substituted from the results in \eqref{TE_stepout_Bn} and \eqref{TE_stepout_Am}. In this plot we took $r_{0}/a=2$ as an example.}
	\label{fig:TEonStepOutBoth}
\end{figure}
Substituting the results from \eqref{L1_found} and \eqref{L2_found} into \eqref{TEStep-out_Bn_balance} now gives the sought answer 
\begin{equation}
    B_{m}=-A_{l}\frac{2aJ_{0}(\nu'_{1l})J_{1}(\nu_{1m}a/r_{0})}{r_{0}\nu_{1m}J^{2}_{0}(\nu_{1m})} \ \ \text{(TE on step-out)}. \label{TE_stepout_Bn}
\end{equation}

To find the coefficients $A_{n}$ we multiply \eqref{TEmatch_Ephi} by $(r\nu'_{1m}/r_{0})J'_{1}(\nu'_{1m}r/r_{0})$, integrate by $\int^{r_{0}}_{0}r dr$ (again, clipping the integral to $\int^{a}_{0}r dr$ on the lhs), then use integration by parts on the second term on the rhs, to yield
\begin{eqnarray}
    &&A_{l}\int^{a}_{0}\frac{\nu'_{1m}r}{r_{0}}J'_{1}(\nu'_{1l}\frac{r}{a})J'_{1}(\nu'_{1m}\frac{r}{r_{0}})dr\nonumber\\
    &&=\sum\limits_{n}A_{n}\int^{r_{0}}_{0}\frac{\nu'_{1m}r}{r_{0}}J'_{1}(\nu'_{1n}\frac{r}{r_{0}})J'_{1}(\nu'_{1m}\frac{r}{r_{0}})dr\nonumber\\
    &&+B_{n}\int^{r_{0}}_{0}J'_{1}(\nu_{1n}\frac{r}{r_{0}})J_{1}(\nu'_{1m}\frac{r}{r_{0}})dr. \label{star1}
\end{eqnarray}

We also multiply \eqref{TEmatch_Er} by $J_{1}(\nu'_{1m}r/r_{0})$ and similarly integrate to obtain
\begin{eqnarray}
    &&A_{l}\int^{a}_{0}\frac{a}{\nu'_{1l} r}J_{1}(\nu'_{1l}\frac{r}{a})J_{1}(\nu'_{1m}\frac{r}{r_{0}})dr\nonumber\\
    && =\sum\limits_{n}\int^{r_{0}}_{0}\frac{A_{n}r_{0}}{\nu'_{1n}r}J_{1}(\nu'_{1n}\frac{r}{r_{0}})J_{1}(\nu'_{1m}\frac{r}{r_{0}})dr\nonumber\\
    &&-B_{n}\int^{r_{0}}_{0}J'_{1}(\nu_{1n}\frac{r}{r_{0}})J_{1}(\nu'_{1m}\frac{r}{r_{0}})dr.\label{start2}
\end{eqnarray}

Adding \eqref{star1} to \eqref{start2} gives
\begin{equation}
    A_{l} (L_{3})=\sum_{n}A_{n} (L_{4}), \label{TEstep-out_An_Balance}
\end{equation}
where $L_{3}$ and $L_{4}$ are the following integrals, whose solution is derived in Appendix~\ref{Appx2}, 
\begin{eqnarray}
L_{3}&=&\int^{a}_{0}dr\left[\frac{\nu'_{1m}r}{r_{0}}J'_{1}(\nu'_{1l}\frac{r}{a})J'_{1}(\nu'_{1m}\frac{r}{r_{0}})\right.\nonumber\\
&& \left. +\frac{a}{\nu'_{1l}r}J_{1}(\nu'_{1l}\frac{r}{a})J_{1}(\nu'_{1m}\frac{r}{r_{0}})\right]\label{L3_first}\\
&=&\frac{r_{0} \nu'_{1m}\nu'_{1l}J_{1}(\nu'_{1l})J'_{1}(\nu'_{1m}a/r_{0})}{(\nu'^{2}_{1l}r_{0}^{2}/a^{2}-\nu'^{2}_{1m})},\label{L3_result}
\end{eqnarray}
\begin{eqnarray}
L_{4}&=&\int^{r_{0}}_{0}dr\left[\frac{\nu'_{1m}r}{r_{0}}J'_{1}(\nu'_{1n}\frac{r}{r_{0}})J'_{1}(\nu'_{1m}\frac{r}{r_{0}})\right.\nonumber\\
&& \left. +\frac{r_{0}}{\nu'_{1n}r}J_{1}(\nu'_{1n}\frac{r}{r_{0}})J_{1}(\nu'_{1m}\frac{r}{r_{0}})\right] \label{L4_first}\\
&=&\delta_{mn}\frac{r_{0}}{2}(\nu'_{1m}-1/\nu'_{1m})J^{2}_{1}(\nu'_{1m}).\label{L4_result}
\end{eqnarray}

Substituting from \eqref{L3_result} and \eqref{L4_result} into \eqref{TEstep-out_An_Balance} now gives
\begin{equation}
    A_{m}=\frac{2A_{l}\nu'_{1l} J_{1}(\nu'_{1l})J'_{1}(\nu'_{1m}\frac{a}{r_{0}})}{(\nu'^{2}_{1l}\frac{r_{0}^{2}}{a^{2}}-\nu'^{2}_{1m})(1-\frac{1}{\nu'^{2}_{1m}})J^{2}_{1}(\nu'_{1m})}\ \ \text{(TE on step-out)}. \label{TE_stepout_Am}
\end{equation}

Using these definitions for $A_{n}$ and $B_{n}$ we can now plot and compare the left-hand and right-hand sides of \eqref{TEmatch_Er} and \eqref{TEmatch_Ephi}, as shown in Fig.~\ref{fig:TEonStepOutBoth}. Good agreement is observed between the original mode and its decomposition.

\subsubsection{\label{TE_step-in} Incident TE on a step-in discontinuity} 

Finally, let us now consider a step-in discontinuity with an incident TE mode (incident from the cavity towards the output waveguide). We write the matched radial electric field equations as
\begin{equation}
    \frac{A_{l}r_{0}}{\nu'_{1l} r}J_{1}(\nu'_{1l}\frac{r}{r_{0}})=\sum\limits_{n}\frac{A_{n}a}{\nu'_{1n}r}J_{1}(\nu'_{1n}\frac{r}{a})-B_{n}J'_{1}(\nu_{1n}\frac{r}{a}),\label{TEmatch_Er2}
\end{equation}
and the matched azimuthal electric field's as
\begin{equation}
    A_{l}J'_{1}(\nu'_{1l}\frac{r}{r_{0}})=\sum\limits_{n}A_{n}J'_{1}(\nu'_{1n}\frac{r}{a})-\frac{B_{n}a}{\nu_{1n}r}J_{1}(\nu_{1n}\frac{r}{a}).\label{TEmatch_Ephi2}
\end{equation}

Notice the symmetry between these equations and equations \eqref{TEmatch_Er} and \eqref{TEmatch_Ephi} for the step-out TE case; the parameters $r_{0}$ and $a$ are interchanged. We proceed with the same approach from last section. To find coefficients $B_{n}$, multiply \eqref{TEmatch_Ephi2} by $J_{1}(\nu_{1m}r/a)$ and integrate across with $\int^{a}_{0}r dr$ to yield
\begin{eqnarray}
    && A_{l}\int^{a}_{0}J'_{1}(\nu'_{1l}\frac{r}{r_{0}})J_{1}(\nu_{1m}\frac{r}{a}) dr\nonumber\\
    && =\sum\limits_{n}A_{n}\int^{a}_{0}J'_{1}(\nu'_{1n}\frac{r}{a})J_{1}(\nu_{1m}\frac{r}{a})dr\nonumber\\
    &&-\int^{a}_{0}\frac{B_{n}a}{\nu_{1n}r}J_{1}(\nu_{1n}\frac{r}{a})J_{1}(\nu_{1m}\frac{r}{a})dr,\label{TEmatch_EphiNew2}
\end{eqnarray}
and we multiply \eqref{TEmatch_Er2} by $\frac{-\nu_{1m}r}{a}J'_{1}(\nu_{1m}r/a)$, integrate, and use integration by parts for the first term on the rhs, to yield
\begin{eqnarray}
    &&\frac{-A_{l}r_{0}\nu_{1m}}{\nu'_{1l}a}\int^{a}_{0}J_{1}(\nu'_{1l}\frac{r}{r_{0}})J'_{1}(\nu_{1m}\frac{r}{a})dr\nonumber\\
    &&=\sum\limits_{n} A_{n}\int^{a}_{0}J_{1}(\nu_{1m}\frac{r}{a})J'_{1}(\nu'_{1n}\frac{r}{a}) dr\nonumber\\
    &&+B_{n}\frac{\nu_{1m}}{a}\int^{a}_{0}rJ'_{1}(\nu_{1n}\frac{r}{a})J'_{1}(\nu_{1m}\frac{r}{a})dr.\label{TEmatch_ErNew3}
\end{eqnarray}

We can find $B_{n}$ now by substracting \eqref{TEmatch_ErNew3} from \eqref{TEmatch_EphiNew2}, which gives
\begin{equation}
    A_{l} (\hat{L}_{1})=-\sum\limits_{n}B_{n}(\hat{L}_{2}), \label{TEStep-out_Bn_balance2}
\end{equation}
where $\hat{L}_{1}$ and $\hat{L}_{2}$ are the following integrals
\begin{eqnarray}
   \hat{L}_{1}&=&\int^{a}_{0} J'_{1}(\nu'_{1l} \frac{r}{r_{0}})J_{1}(\nu_{1m}\frac{r}{a})dr \nonumber\\
   &&  +\frac{a\nu_{1m}}{a \nu'_{1l}}\int^{a}_{0}J_{1}(\nu'_{1l}\frac{r}{r_{0}})J'_{1}(\nu_{1m}\frac{r}{a})dr\label{L1hat_first}\\
    &=&\frac{r_{0}}{\nu'_{1l}}\left[  J_{1}(\nu'_{1l}\frac{r}{r_{0}})J_{1}(\nu_{1m}\frac{r}{a}) \right]^{a}_{0}=0, \label{L1found2}\\
    \hat{L}_{2}&=&\sum\limits_{n}-B_{n}\int^{a}_{0} \frac{a}{\nu_{1n}r}J_{1}(\nu_{1n}\frac{r}{a})J_{1}(\nu_{1m}\frac{r}{a})dr\nonumber\\
    &&+\frac{\nu_{1m}}{a}\int^{a}_{0}rJ'_{1}(\nu_{1n}\frac{r}{a})J'_{1}(\nu_{1m}\frac{r}{a})dr \label{L2hat_first}\\
    &\equiv&\frac{\nu_{1m}}{a}I_{1}=\delta_{mn}\frac{a\nu_{1m}}{2}J^{2}_{0}(\nu_{1m}), \label{L2found2}
\end{eqnarray}
where $I_{1}$ has been calculated earlier in \eqref{I1_result}. Upon substituting into \eqref{TEStep-out_Bn_balance2}, we find that 
\begin{equation}
    B_{m}=0 \ \ \text{(TE on step-in)}.\label{TE_stepin_Bn}
\end{equation}
\begin{figure}
	\includegraphics[width=0.8\columnwidth]{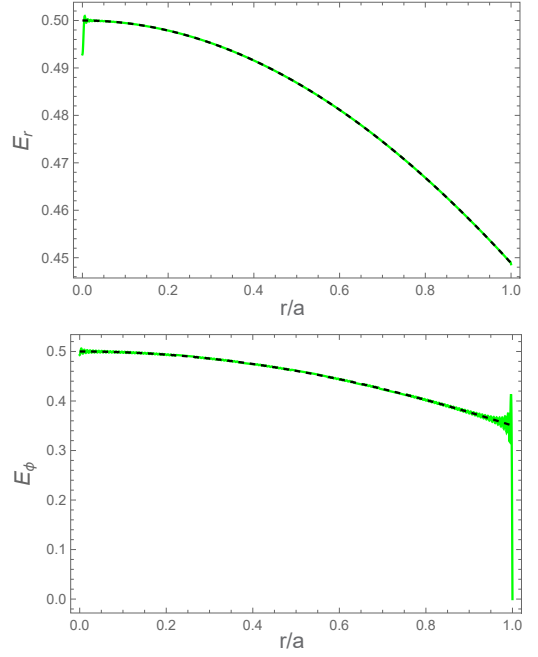}
	\caption{Comparison of the right-hand sides (green color line) with the left-hand sides (dashed black line) of equation \eqref{TEmatch_Er2} for $E_{r}$ and equation \eqref{TEmatch_Ephi2} for $E_{\phi}$, with the upper limit of the sums truncated at 300 terms and the $(A_{n},B_{n})$ values substituted from the results in \eqref{TE_stepin_Bn} and \eqref{TE_stepin_Am}. In this plot we took $r_{0}/a=2$ as an example.}
	\label{fig:TEonStepInBoth}
\end{figure}
Notice the symmetry between this TE step-in case (giving $B_{m}=0$) and the TM step-out case (giving $A_{m}=0$) in Subsection~\ref{TM_step-out}.

Table~\ref{tab1} summarizes the above results for all four combinations of incident-mode type and discontinuity type, from Subsection~\ref{TM_step-out} to \ref{TE_step-in}.
\begin{table*}
\begin{center}
\caption{Summary of results for the forward-scatter expansion coefficients $(A_{n},B_{n})$, for a TE/TM mode incident on a step-out/in discontinuities in the oversized iris-line cell under consideration. Notice the symmetry of zeros in the table's diagonal.}
\label{tab1}
\begin{tabular}{ |c|c|c| } 
\hline
 &  \multicolumn{2}{c|}{Output TE$_{n}$/TM$_{n}$ mode coefficients ($A_{n}$ for TE$_{n}$ and $B_{n}$ for TM$_{n}$)} \\ 
\hline
 & \textbf{Step-out} & \textbf{Step-in} \\
\hline\hline
\multirow{5}{4em}{\textbf{TM$_{l}$} \linebreak incident mode} & &\\
&  $A_{n}=0 $ & $A_{n}=-B_{l}\frac{2 r_{0}J_{1}(\nu_{1l}a/r_{0})}{a \nu_{1l}(\nu'_{1n}-1/\nu'_{1n})J_{1}(\nu'_{1n})}$ \\ 
& &\\
& $B_{n}=-B_{l}\frac{2a\nu_{1n}J_{0}(\nu_{1l})J_{1}\left( \nu_{1n}\frac{a}{r_{0}}\right)}{r_{0}^3 (\nu_{1l}^2/ a^2-\nu_{1n}^2 /r_{0}^2)J'^{2}_{1}\left( \nu_{1n}\right)}$ & $B_{n}=-B_{l}\frac{2a\nu_{1l}J_{0}(\nu_{1n})J_{1}(\nu_{1l}a/r_{0})}{r_{0}(\nu^{2}_{1n}-\nu^{2}_{1l}a^{2}/r_{0}^{2})J'^{2}_{1}(\nu_{1n})}$ \\ 
& &\\
\hline
\multirow{5}{4em}{\textbf{TE$_{l}$}\linebreak incident mode} & &\\
& $A_{n}=A_{l}\frac{2\nu'_{1l} J_{1}(\nu'_{1l})J'_{1}(\nu'_{1n}\frac{a}{r_{0}})}{(\nu'^{2}_{1l}r_{0}^{2}/a^{2}-\nu'^{2}_{1n})(1-1/\nu'^{2}_{1n})J^{2}_{1}(\nu'_{1n})}$ & $A_{n}=A_{l}\frac{2r_{0}^{2}\nu'_{1n} J'_{1}(\nu'_{1l}\frac{a}{r_{0}})}{a^{2}(\nu'^{2}_{1n}r_{0}^{2}/a^{2}-\nu'^{2}_{1l})(1-1/\nu'^{2}_{1n})J_{1}(\nu'_{1n})}$ \\ 
& &\\
& $B_{n}=-A_{l}\frac{2aJ_{0}(\nu'_{1l})J_{1}(\nu_{1n}a/r_{0})}{r_{0}\nu_{1n}J^{2}_{0}(\nu_{1n})}$ & $B_{n}=0$ \\ 
& &\\
\hline
\end{tabular}
\end{center}
\end{table*}

To find the coefficients $A_{n}$ we multiply \eqref{TEmatch_Ephi2} by $(r\nu'_{1m}/a)J'_{1}(\nu'_{1m}r/a)$, integrate across with $\int^{a}_{0}rdr$, and use integration by parts on the second term on the rhs, to yield
\begin{eqnarray}
    &&A_{l}\int^{a}_{0}\frac{\nu'_{1m}r}{a}J'_{1}(\nu'_{1l}\frac{r}{r_{0}})J'_{1}(\nu'_{1m}\frac{r}{a})dr\nonumber\\
    &&=\sum\limits_{n}A_{n}\int^{a}_{0}\frac{\nu'_{1m}r}{a}J'_{1}(\nu'_{1n}\frac{r}{a})J'_{1}(\nu'_{1m}\frac{r}{a})dr\nonumber\\
    &&+B_{n}\int^{a}_{0}J'_{1}(\nu_{1n}\frac{r}{a})J_{1}(\nu'_{1m}\frac{r}{a})dr. \label{star1b}
\end{eqnarray}

We also multiply \eqref{TEmatch_Er2} by $J_{1}(\nu'_{1m}r/a)$ and similarly integrate across to obtain
\begin{eqnarray}
    &&A_{l}\int^{a}_{0}\frac{r_{0}}{\nu'_{1l} r}J_{1}(\nu'_{1l}\frac{r}{r_{0}})J_{1}(\nu'_{1m}\frac{r}{a})dr\nonumber\\
    && =\sum\limits_{n}\int^{a}_{0}\frac{A_{n}a}{\nu'_{1n}r}J_{1}(\nu'_{1n}\frac{r}{a})J_{1}(\nu'_{1m}\frac{r}{a})dr\nonumber\\
    &&-B_{n}\int^{a}_{0}J'_{1}(\nu_{1n}\frac{r}{a})J_{1}(\nu'_{1m}\frac{r}{a})dr.\label{start2b}
\end{eqnarray}

Adding \eqref{star1b} to \eqref{start2b} yields
\begin{equation}\label{TEstep-out_An_Balance2}
    A_{l}(\hat{L}_{3})=\sum_{n}A_{n}(\hat{L}_{4}),
\end{equation}
where $\hat{L}_{3}$ and $\hat{L}_{4}$ are the following integrals, whose solution is derived in Appendix~\ref{Appx2},
\begin{eqnarray}
\hat{L}_{3}&=& \int^{a}_{0}dr \frac{\nu'_{1m}r}{a}J'_{1}(\nu'_{1l}\frac{r}{r_{0}})J'_{1}(\nu'_{1m}\frac{r}{a})\nonumber\\
&&+\int^{a}_{0}dr\frac{a}{\nu'_{1l}r}J_{1}(\nu'_{1l}\frac{r}{r_{0}})J_{1}(\nu'_{1m}\frac{r}{a})\label{L3hat_first}\\
&=&\frac{a r_{0}^{2} \nu'^{2}_{1m}J_{1}(\nu'_{1m})J'_{1}(\nu'_{1l}a/r_{0})}{a^{2}(\nu'^{2}_{1m}r_{0}^{2}/a^{2}-\nu'^{2}_{1l})}\label{L3hat_result}\\
\hat{L}_{4}&=&\sum_{n}A_{n}\int^{a}_{0} dr\left[ \frac{\nu'_{1m}r}{a}J'_{1}(\nu'_{1n}\frac{r}{a})J'_{1}(\nu'_{1m}\frac{r}{a})\right.\nonumber\\
&& \left. +\frac{a}{\nu'_{1n}r}J_{1}(\nu'_{1n}\frac{r}{a})J_{1}(\nu'_{1m}\frac{r}{a})\right] \label{L4hat_first}\\
&=&\delta_{mn}\frac{a}{2}(\nu'_{1m}-1/\nu'_{1m})J^{2}_{1}(\nu'_{1m}). \label{L4hat_result}
\end{eqnarray}

Substituting from \eqref{L3hat_result} and \eqref{L4hat_result} into \eqref{TEstep-out_An_Balance2} gives
\begin{equation}
    A_{m}=A_{l}\frac{2r_{0}^{2}\nu'_{1m} J'_{1}(\nu'_{1l}\frac{a}{r_{0}})}{a^{2}(\nu'^{2}_{1m}r_{0}^{2}/a^{2}-\nu'^{2}_{1l})(1-1/\nu'^{2}_{1m})J_{1}(\nu'_{1m})}\ \ \text{(TE on step-in)} \label{TE_stepin_Am}
\end{equation}

Using these definitions for $A_{n}$ and $B_{n}$ we can now plot and compared the left-hand and right-hand sides of \eqref{TEmatch_Er2} and \eqref{TEmatch_Ephi2}, as shown in Figure~\ref{fig:TEonStepInBoth}. Good agreement is observed between the original mode and its decomposition.

\begin{figure}
	\includegraphics[width=0.8\columnwidth]{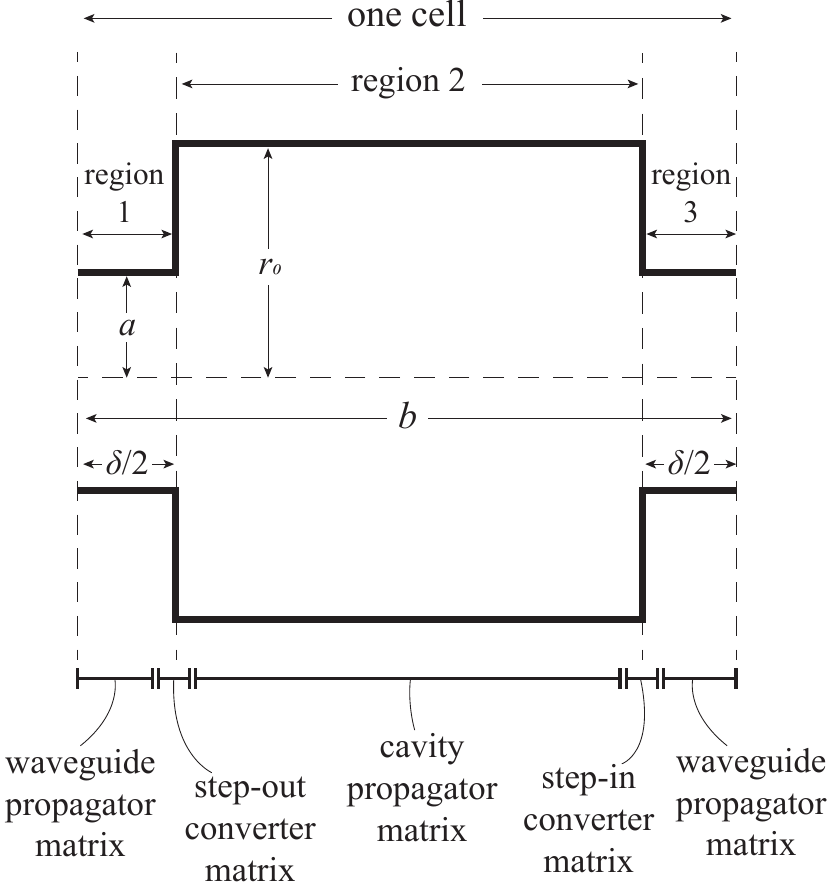}
	\caption{A sketch illustrating how the different sections of the cell are modelled using different transmission matrices in the numerical analysis.}
	\label{fig:MatrixFig}
\end{figure}

\section{\label{sec: implementation} Implementation and numerical examples}

The theoretical forward-scatter analysis developed in Section~\ref{sec:formalism} can now be numerically implemented on a computer to predict the performance of practical iris-line examples. The iris line will be represented as a series of connected cells, the model of each of which is a transmission matrix that captures the modal decompositions summarized in Table~\ref{tab1}. The periodic iris line is therefore conveniently represented by matrix multiplication or raising the matrix to power $M$, where $M$ is the number of cells in a practical iris line. We can look at the field profiles at the output of each cell, to examine the transient regime, mode launching purity and how an excited mode settles gradually along the line. It is clear that anomalies or defects in the geometry can also be incorporated by adapting this approach to include some cells that are different from the remaining (periodic) ones. In this section, we discuss implementation aspects and illustrate the theory using numerical examples. We also compare the diffraction loss behavior obtained from this method with that obtained from the mode-matching method \cite{NajiTHz1}, demonstrating good agreement, but faster computations.

\subsection{\label{sec: code} Transmission matrix representation}

Fig.~\ref{fig:MatrixFig} illustrates a conceptual picture of how we may represent the cell matrix based on the decompositions found in the field equations \eqref{ArhoTE}--\eqref{HzTMc} and Table~\ref{tab1}. The signal from the source is represented by an input vector (a mix of TE and TM modes) and each cell of the iris line is represented by five matrices:  a propagator matrix for the input waveguide (region 1), a converter matrix for the step-out transition, a propagator matrix for the cavity (region 2), a converter matrix for the step-in transition, and finally a propagator matrix for the output waveguide (region 3). The output of the cell is another vector (mix of TE and TM modes) that is then inputted to the following cell, and so forth. We may also have the freedom of truncating the modal sums in each region arbitrarily, where $N_{1},N_{2}$ and $N_{3}$ denote the number of terms included in each of the TE sums and TM sums, for regions 1, 2 and 3, respectively (see Fig.~\ref{fig:MatrixFig}).

Using block matrices, the input/output vectors for the $E_{r}$ field, for example, can be represented by the column matrices
\begin{equation} \label{in-out-vectors}
    \mathsf{E}_{r,\text{in}}
    \equiv
    \left(\begin{matrix}
    A_{1} \\
    A_{2} \\
    \vdots \\
    A_{N_{1}}\\
    \hline
    B_{1}\\
    B_{2}\\
    \vdots\\
    B_{N_{1}}\\
    \end{matrix}
    \right)
    \begin{matrix}
    \uparrow \\
   \footnotesize{\text{TE}_{N_{1}}} \\
    \downarrow \\
    \\
    \\
    \uparrow \\
    \footnotesize{\text{TM}_{N_{1}}} \\
    \downarrow \\
    \end{matrix}
   ,  \ \ 
    \mathsf{E}_{r,\text{out}}
    \equiv
    \left(\begin{matrix}
    \tilde{A}_{1} \\
    \tilde{A}_{2} \\
    \vdots \\
    \tilde{A}_{N_{3}}\\
    \hline
    \tilde{B}_{1}\\
    \tilde{B}_{2}\\
    \vdots\\
    \tilde{B}_{N_{3}}\\
    \end{matrix}
    \right)
    \begin{matrix}
    \uparrow \\
    \footnotesize{\tilde{\text{TE}}_{N_{3}}} \\
    \downarrow \\
    \\
    \\
    \uparrow \\
    \footnotesize{\tilde{\text{TM}}_{N_{3}}} \\
    \downarrow \\
    \end{matrix}
    ,
\end{equation}
where we use San Serif symbols, such as $\mathsf{E}$, to denote matrices. We divide each vector into an upper half that contains the amplitudes of its TE modes (the $A_{n}$ coefficients) and a lower half that contains the amplitudes of its TM modes (the $B_{n}$ coefficients). Uppercase $N$ symbols denote the number of terms taken for TE or TM in each region (e.g.~$\text{TE}_{N_{1}}$ denotes having as many as $N_{1}$ modes of type TE). The tilde superscript is used above some of the symbols of the output vector to distinguish them from those of the input vector. Similar columns are used for the other field components, such as $E_{\phi}, H_{r}$, etc. 

For the straight sections along the $z$ direction, we know that the mode compositions are not altered, but merely `propagated' in phase as they travel down the line. This can be readily inferred from the field exponents given by equation \eqref{ArhoTE}--\eqref{HzTE} and \eqref{ArhoTM}--\eqref{HzTM} for the waveguide sections, for example, which are $e^{-iz\frac{\nu'^{2}_{1n}}{2ka^{2}}+ikz}$ (for TE modes) or $e^{-iz\frac{\nu^{2}_{1n}}{2ka^{2}}+ikz}$ (for TM modes); with similar exponents for the cavity section (replace $a \leftrightarrow r_{0}$). The propagator sections for the input waveguide or cavity, for example, can be represented by the diagonal matrices
\begin{equation} \label{propagator1}
    \mathsf{P}_{1}
    \equiv
    \left(\begin{matrix}
    e^{-iz\frac{\nu'^{2}_{11}}{2ka^{2}}+ikz} & 0 & \cdots & 0 \\
    0& e^{-iz\frac{\nu'^{2}_{12}}{2ka^{2}}+ikz} & \cdots & 0\\
    \vdots & \vdots & \ddots & \\
    0& 0 & \cdots & e^{-iz\frac{\nu'^{2}_{1N_{1}}}{2ka^{2}}+ikz}\\
    \hline
   e^{-iz\frac{\nu^{2}_{11}}{2ka^{2}}+ikz} & 0 & \cdots & 0 \\
    0& e^{-iz\frac{\nu^{2}_{12}}{2ka^{2}}+ikz} & \cdots & 0\\
    \vdots & \vdots & \ddots & \\
    0& 0 & \cdots & e^{-iz\frac{\nu^{2}_{1N_{1}}}{2ka^{2}}+ikz}\\
    \end{matrix}
    \right), 
\end{equation}
\begin{equation} \label{propagator2}
    \mathsf{P}_{2}
    \equiv
    \left(\begin{matrix}
    e^{-iz\frac{\nu'^{2}_{11}}{2kr_{0}^{2}}+ikz} & 0 & \cdots & 0 \\
    0& e^{-iz\frac{\nu'^{2}_{12}}{2kr_{0}^{2}}+ikz} & \cdots & 0\\
    \vdots & \vdots & \ddots & \\
    0& 0 & \cdots & e^{-iz\frac{\nu'^{2}_{1N_{2}}}{2kr_{0}^{2}}+ikz}\\
    \hline
   e^{-iz\frac{\nu^{2}_{11}}{2kr_{0}^{2}}+ikz} & 0 & \cdots & 0 \\
    0& e^{-iz\frac{\nu^{2}_{12}}{2kr_{0}^{2}}+ikz} & \cdots & 0\\
    \vdots & \vdots & \ddots & \\
    0& 0 & \cdots & e^{-iz\frac{\nu^{2}_{1N_{2}}}{2kr_{0}^{2}}+ikz}\\
    \end{matrix}
    \right).
\end{equation}

Multplying the $E_{r}$ field's input column, $\mathsf{E}_{r}$, from \eqref{in-out-vectors} by the $\mathsf{P}_{1}$ matrix from \eqref{propagator1}, for example, will give us the state of the field composition after traveling along the input waveguide section of the iris line for a distance $z$. 

We use `converter' matrices to represent the step-out/in transitions in a similar way. For example, the step-out transition from the input waveguide section (region 1) to the cavity section (region 2) is given by
\begin{equation} \label{converter-matrix}
   \mathsf{S}_{\text{step-out}}
    {\equiv}
   \left(\begin{array}{ccc|ccc}
    A^{\text{TE}}_{1,1}  & \cdots & A^{\text{TE}}_{1,N_{2}} & B^{\text{TE}}_{1,1}  & \cdots & B^{\text{TE}}_{1,N_{2}} \\
    A^{\text{TE}}_{2,1}  & \cdots & A^{\text{TE}}_{2,N_{2}} & B^{\text{TE}}_{2,1}  & \cdots & B^{\text{TE}}_{2,N_{2}} \\
    \vdots    &        &  \vdots     & \vdots    &        & \vdots \\
    A^{\text{TE}}_{N_{1},1}  & \cdots & A^{\text{TE}}_{N_{1},N_{2}} & B^{\text{TE}}_{N_{1},1}  & \cdots & B^{\text{TE}}_{N_{1},N_{2}} \\
   &  & & &  &  \\
    \hline &  & & &  & \\
    A^{\text{TM}}_{1,1}  & \cdots & A^{\text{TM}}_{1,N_{2}} & B^{\text{TM}}_{1,1}  & \cdots & B^{\text{TM}}_{1,N_{2}} \\
    A^{\text{TM}}_{2,1}  & \cdots & A^{\text{TM}}_{2,N_{2}} & B^{\text{TM}}_{2,1}  & \cdots & B^{\text{TM}}_{2,N_{2}} \\
    \vdots    &        &  \vdots     & \vdots   &        & \vdots \\
    A^{\text{TM}}_{N_{1},1}  & \cdots & A^{\text{TM}}_{N_{1},N_{n}} & B^{\text{TM}}_{N_{1},1}  & \cdots & B^{\text{TM}}_{N_{1},N_{2}} \\
    \end{array}
    \right),
\end{equation}
where, $A^{\text{TE}}_{ln}$ denotes the $A_{n}$ coefficients obtained from Table~\ref{tab1} when the incident mode is the $l$-th TE mode, and so forth. Thus, the upper half of matrix \eqref{converter-matrix} is the decomposition matrix when the TE modes are incident (we have $N_{1}$ of such modes). Each row among the top half, call it the $l$-th row, therefore represents the $(A_{n},B_{n})$ values when the $l$-th TE mode is incident. For the lower part, TM modes are incident, and we interpret the rows in the same way.  Within each row, the first $N_{2}$ terms represent the TE modes scattered forward from the transition into the cavity, whereas the second $N_{2}$ terms are the TM modes scattered into the cavity. Note that terms in each entire column of the matrix actually represent the same cavity mode (same mode at the output of the transition), but each coming from a different waveguide-mode contribution (inputs of the transition). Matrix multiplication between an input vector and the converter matrix like \eqref{converter-matrix} will cause each of these columns to be weighted appropriately by the input mode amplitudes and then collapsed (summed up) into one number, which represents the corresponding mode coming out of the transition on the cavity side.

It is now clear that an entire cell can be represented by the multiplication of the matrices introduced above. For example, the $E_{r}$ field at the output of the cell is given by
\begin{equation}
    \mathsf{E}^{\mathsf{T}}_{r,\text{out}}=\mathsf{E}^{\mathsf{T}}_{r,\text{in}}\cdot \underbrace{\mathsf{P}_{1} \cdot \mathsf{S}_{\text{step-out}} \cdot \mathsf{P}_{2} \cdot \mathsf{S}_{\text{step-out}} \cdot \mathsf{P}_{3}}_{\equiv \ \mathsf{C}},
\end{equation}
where $\mathsf{C}$ denotes the equivalent cell matrix and the $\mathsf{T}$ superscript denotes taking the transpose of a matrix. For a periodic iris line that holds $M$ such cells, we can simply represent the entire iris line by the power law $\mathsf{L}=\mathsf{C}^{M}$, where $\mathsf{L}$ represents the matrix for the entire line.  For an iris line that has, for example, 50 periodic cells, followed by a mechanically deformed cell, then by another 20 periodic cells, we can find the equivalent matrix for the line by $\mathsf{L}=\mathsf{C}^{50}\cdot \mathsf{\hat{C}}\cdot\mathsf{C}^{20}$, where $\mathsf{C}$ represents any of the periodic cells and $\mathsf{\hat{C}}$ the deformed cell. 
 
Matrix multiplication or exponentiation operators can be efficiently computed on modern computers using standard tools, such as Wolfram Mathematica. Indeed, the developed model has proved to be typically an order-of-magnitude faster than the mode-matching approach reported in \cite{NajiTHz1}, as in the latter we had to solve the numerically challenging problem of finding the zeros of a complex determinant for the eigenvalue problem. For the same iris line setup (for example see Subsection~\ref{sec:example3} below) and using the same computational tool, when the mode-matching technique took several hours to find a solution point (e.g.~see Scale-3 in Table~II in reference~\cite{NajiTHz1}), the current method typically took only a few minutes on the same computer.

\subsection{\label{sec:examples} Numerical examples}

In the following examples, we apply the developed theoretical model to the same iris line structure under different excitation scenarios. To facilitate the comparison with the diffraction loss results found previously by the perturbative mode-matching method \cite{NajiTHz1} (for arbitrary $\delta$) and the Vainstein-boundary condition method \cite{DESY} (for $\delta\rightarrow 0$), we choose an iris-line example with an iris radius of 5.5~cm, a period of 33.3~cm, length of 150~m (approximately equivalent to 450 cells and 451 irises) and a frequency of 3~THz (0.1~mm wavelength). In these examples all the modal sums were truncated to $N_{1}=N_{2}=N_{3}=500$ terms, which was more than sufficient for an accuracy better than 0.1\% in all the diffraction loss results. The diffraction loss is calculated from the transmission matrix of the full line ($\mathsf{L}$), by comparing the total power (from all modes) at the output with the inputted power, in the absence of ohmic losses (see Section~\ref{sec:OtherConsiderations} for more details on the ohmic loss).  The iris thickness is taken at the nominal value of $\delta=2$~mm, except when it is being varied, and the chamber radius for the present closed model is chosen arbitrarily as twice the iris's radius (see \cite{NajiTHz1, Palmer, Lawson1,BaneWilson1} for a discussion on the chamber radius and diffraction scale approximations). 

To avoid having overcrowded transient plots, we plot transient samples at fixed intervals (every 50 irises) along the line. We also plot the field amplitudes with and without normalization with respect to their amplitude on axis ($r=0$). The normalized plots help in observing how the profile shape (as a function of $r$) of the propagating mode is evolving along the line. For the sake of demonstration, we demonstrate the transients in the $E_{r}$ field across all the examples; similar plots can be calculated for the other field components.

\begin{figure}
	\includegraphics[width=\columnwidth]{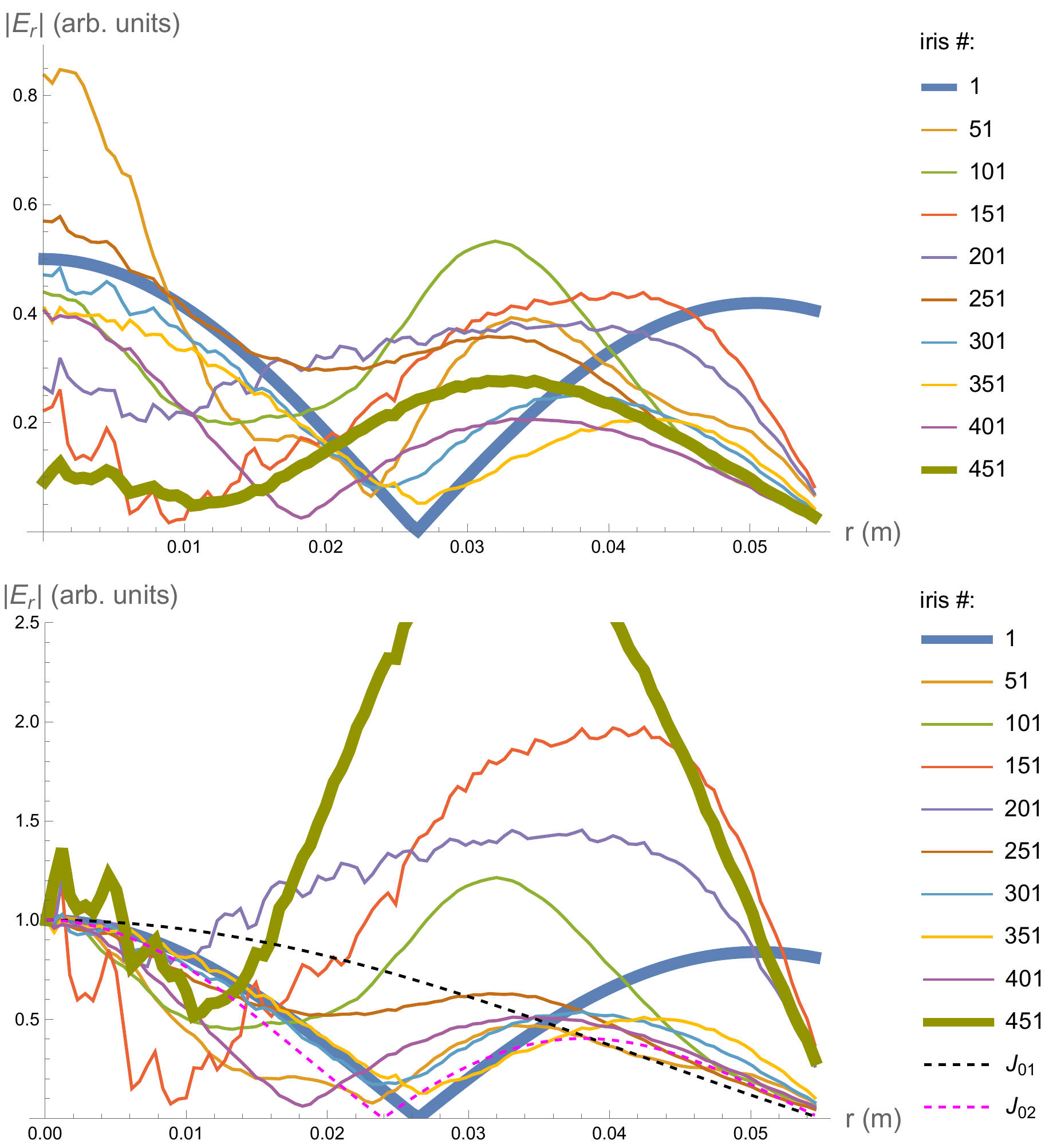}
	\caption{Transients along the iris line (sampled at the indicated irises) for the TM$_{11}$ excitation case. The plots show the magnitude of the field component $E_{r}$, without (top) and with (bottom) normalization  with respect to the amplitude on the axis $r=0$. Heavy trace lines in the plot highlight the field profiles at the first iris and last iris of the line, while dashed lines highlight the first two eigenmodes of the iris line (denoted $J_{01}, J_{02}$).}
	\label{fig:transientTM11}
\end{figure}
\begin{figure}
	\includegraphics[width=\columnwidth]{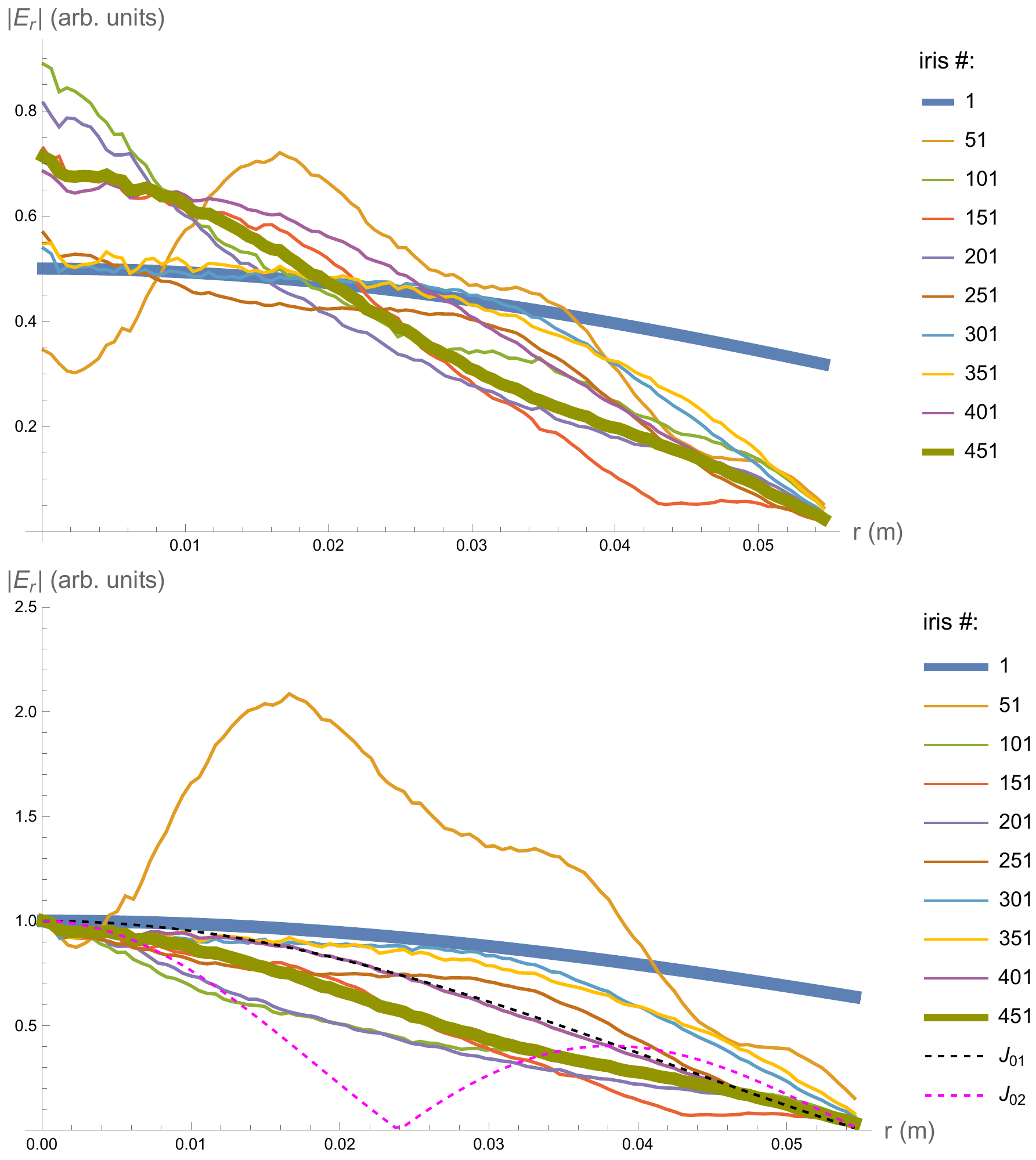}
	\caption{Transients along the iris line (sampled at the indicated irises) for the TE$_{11}$ excitation case. The plots show the magnitude of the field component $E_{r}$, without (top) and with (bottom) normalization with respect to the amplitude on the axis $r=0$. Heavy trace lines in the plot highlight the field profiles at the first iris and last iris of the line, while dashed lines highlight the first two eigenmodes of the iris line (denoted $J_{01}, J_{02}$).}
	\label{fig:transientTE11}
\end{figure}

\subsubsection{\label{sec:example2} An iris line with a TM$_{11}$ source }

If a TM$_{11}$ mode is launched from the source inside the input waveguide (region 1), one can expect a mismatch with the dominant eignemodes supported by the iris line. This is, indeed, demonstrated in Fig.~\ref{fig:transientTM11}, where the transient field profiles are sampled at fixed intervals along the line. The instantaneous amplitude and phase of the transients will generally change from iris to next, but their profile will evolve to eventually settle, after a sufficient distance, on one of the eigenmodes of the line or a linear combination thereof. Ideally, we aim to minimize mode-competition and maximize mode purity by matching the inputted mode to the dominant (least lossy) eigenmode, which is approximated \cite{NajiTHz1,DESY} by the function $J_{0}(2.4r/a)$, and which will be the main surviving mode in the case of a long iris line. This eigenmode is plotted in Fig.~\ref{fig:transientTM11}, alongside the second eigenmode approximated by $J_{0}(5.52 r/a)$, for reference. The settlement of the transient can in general take a short or long distance from the input, depending on the profile of the source signal. In this case, the input is not well matched to any of the leading eigemodes and, for a total length of 150~m, the wave arrives at the final iris with a overall diffraction loss ($L$) of 53.5\%, which is relatively high.

\subsubsection{\label{sec:example1} An iris line with a TE$_{11}$ source }

As in the case of a TM$_{11}$ source, launching a TE$_{11}$ mode in the input waveguide will lead to a mismatch between the excited mode and the eigenmodes supported by the iris line, but to a lesser extent, since the TE$_{11}$ mode is closer in profile to the dominant eigenmode, $J_{0}(2.4r/a)$. Fig.~\ref{fig:transientTE11} demonstrated this fact, by showing the sampled transients along the line. At the last iris, the total accrued loss ($L$) is 21.9\%, which is much better than that for the TM$_{11}$ mode.

\subsubsection{\label{sec:example3} An iris line with a well-matched Bessel source }

\begin{figure}
	\includegraphics[width=\columnwidth]{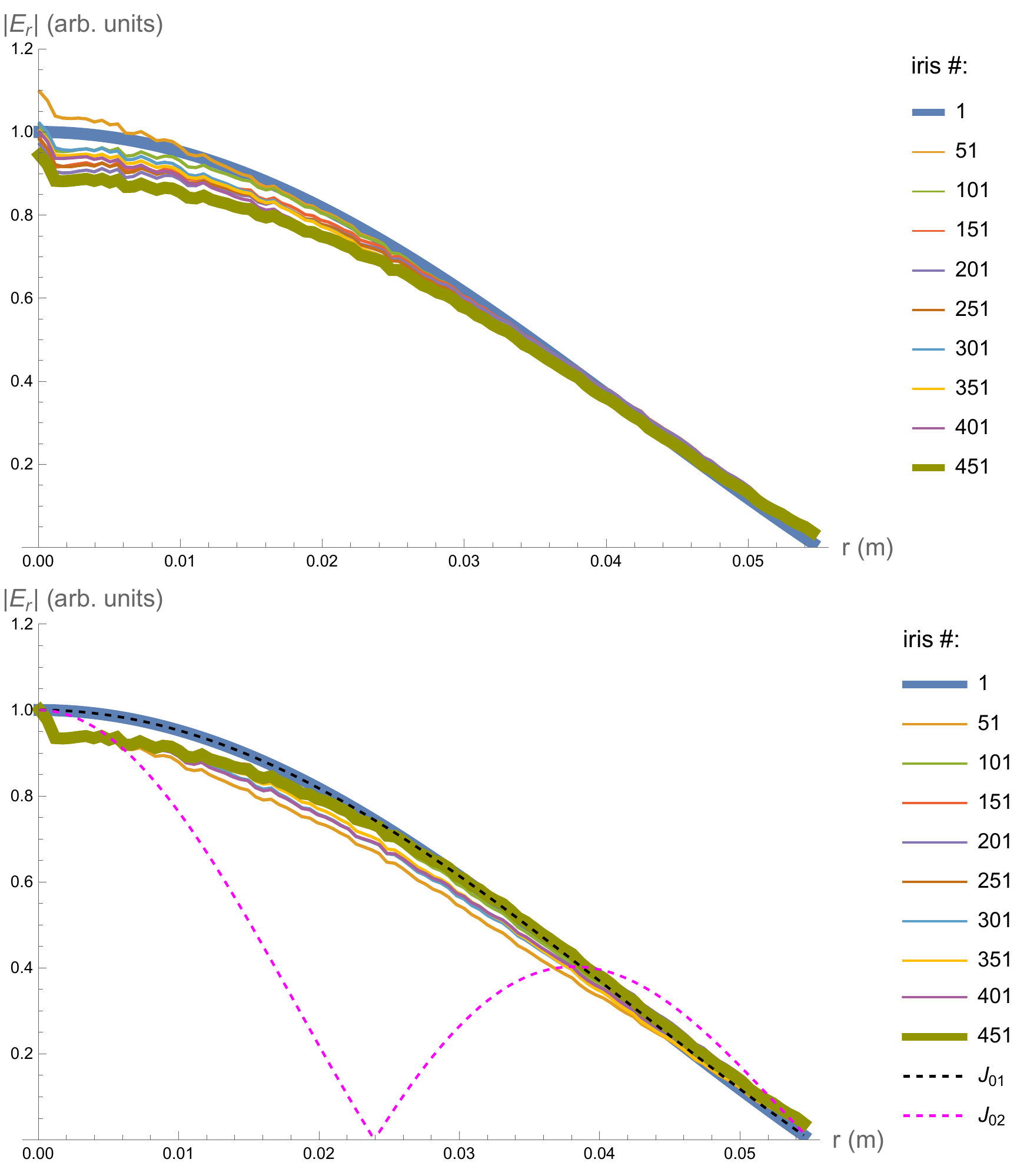}
	\caption{Transients along (sampled at the indicated irises) the iris line for the matched $J_{0}(2.4r/a)$ source case. The plots show the magnitude of the field component $E_{r}$, without (top) and with (bottom) normalization with respect to the amplitude on the axis $r=0$. Heavy trace lines in the plot highlight the field profiles at the first iris and last iris of the line, while dashed lines highlight the first two eigenmodes of the iris line (denoted $J_{01}, J_{02}$).}
	\label{fig:transientJ0}
\end{figure}
\begin{figure}
	\includegraphics[width=\columnwidth]{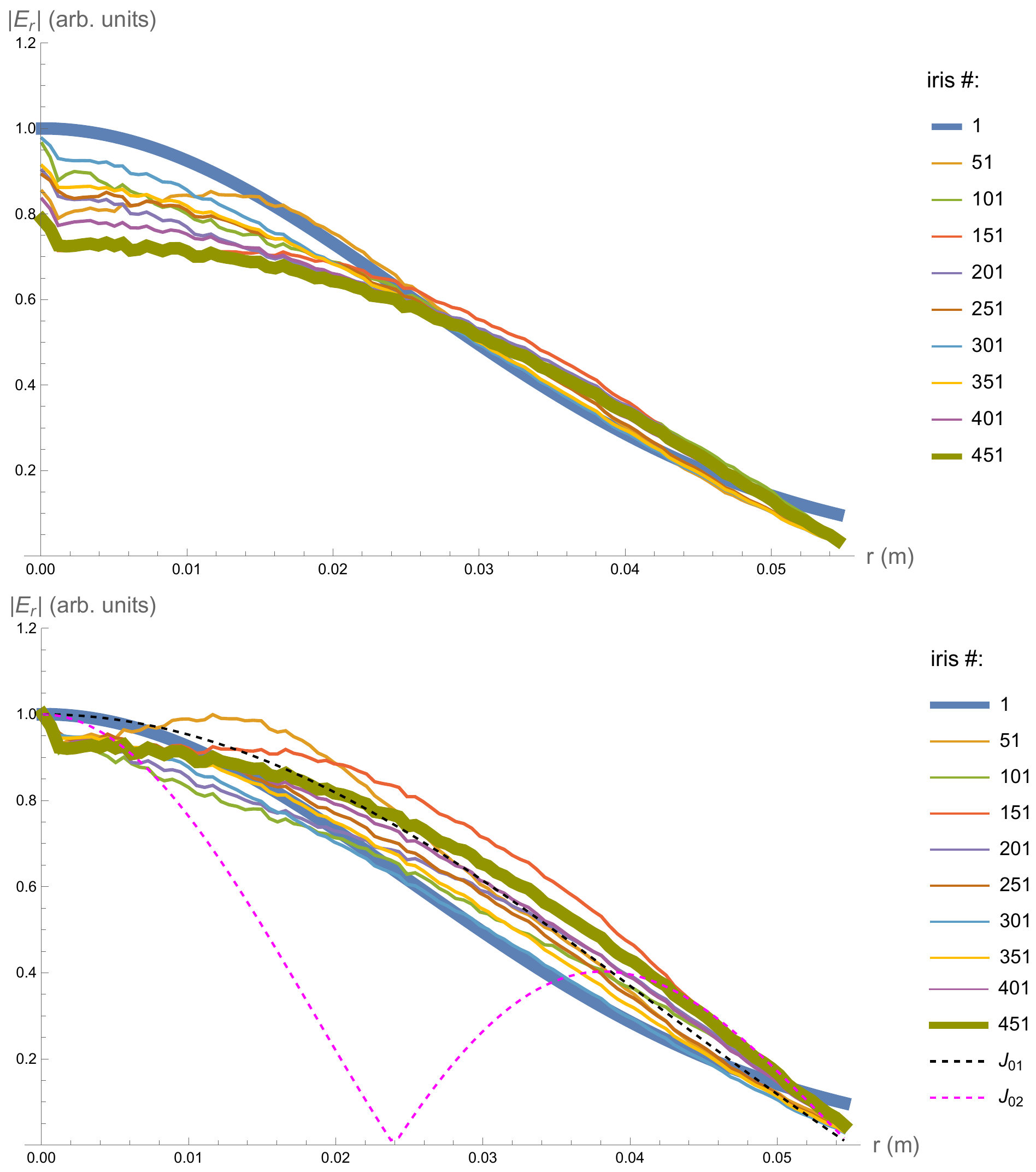}
	\caption{Transients  along the iris line (sampled at the indicated irises) for the backed-off Gaussian excitation case. The plots show the magnitude of the field component $E_{r}$, without (top) and with (bottom) normalization with respect to the amplitude on the axis $r=0$. Heavy trace lines in the plot highlight the field profiles at the first iris and last iris of the line, while dashed lines highlight the first two eigenmodes of the iris line (denoted $J_{01}, J_{02}$).}
	\label{fig:transientGaussian}
\end{figure}

When the line is excited by a mode that is well-matched to its dominant mode, approximated by $J_{0}(2.4r/a)$, one would expect a minimal transient regime, since the launched field's profile does not need to evolve significantly to meet the dominant eigenmode's profile. Indeed, this is equivalent to the eigensolution scenario used by the mode-matching method \cite{NajiTHz1} for an infinitely-long line, where the mode is assumed to have already settled in the steady state. Fig.~\ref{fig:transientJ0} shows the sampled transients for this case. As expected, this excitation type results in the least accrued diffraction loss over the line, with $L=13.6$\%. This least-lossy scenario requires a Bessel mode source, which may not be readily available in practice compared to other more-common sources. Sources with a Gaussian profile, for example, may be easier to find in practice and are considered below. Note that a linear THz undulator may also be approximated as a Gaussian source \cite{NajiTHz1,FELbook}.

\subsubsection{\label{sec:example4} An iris line with a nearly-matched Gaussian source}

Consider now a Gaussian beam source whose beam-waist is aligned with the input plane of the cell, so as to have flat wavefronts (no curvature) at the waveguide's input iris. The skirt of the Gaussian profile, $e^{-r^{2}/w^{2}}$, can be controlled by adjusting $w$ relative to the iris radius $a$, where $w$ is the $1/e^{2}$ width of the Gaussian. An optimal overlap between the skirts of the Gaussian function and the $J_{0}(2.4r/a)$ Bessel function may be obtained by backing off from the iris's full width (i.e.~putting $w<a$). This maximizes the power transfer (projection) between the two modes, leading to a shorter transient regime and a lower diffraction loss overall. Fig.~\ref{fig:transientGaussian} shows the sampled transients corresponding to the value of  $w\approx0.65a$, which results in a minimal (`nearly-matched') diffraction loss on the line. The total loss accrued in this case is $L=14.3$\%, which is a little higher than the ideal case of the Bessel source given in the previous example.  It is worth noting that a Gaussian beam that fills the iris with $w=a$ would have given a total loss of $L=18.8\%$ in comparison.

\begin{table}[t]
\caption{\label{tab:results}
Comparison of the diffraction loss ($L$) results obtained by different models for a matched Bessel source (the dominant eigenmode) and a backed-off Gaussian source, as we vary the thickness $\delta$ of the thin screens. For the well-matched input (or eigenmode), which has a profile ${\sim} J_{0}(2.4 r/a)$, the comparison is between the Vainstein-based model (for $\delta=0$) \cite{DESY}, the mode-matching model (for any $\delta$) \cite{NajiTHz1}, and the present forward-scatter model (abbreviated as `Fwd-Sctr'). For the backed-off Gaussian input, which has a profile ${\sim}\text{exp}[-r^{2}/(0.65 a)^{2}]$, the comparison is only meaningful with the matched Bessel mode using the present model. All modes are assumed linearly polarized and $\delta$ is varied over the indicative values used in \cite{NajiTHz1}. A frequency of 3~THz is assumed, with iris-line dimensions: $a=5.5$~cm, $b=33.3$~cm and total length of 150~m. For this calculation, the chamber radius was taken arbitrarily as $r_{0}\approx 2a$ in the present model (the other models assume open chambers).}
\begin{ruledtabular}
\begin{tabular}{c|c|c|c|c}
  & \multicolumn{3}{c|}{Eigenmode or well-matched source } & Gaussian source \\
 \hline
$\delta$  & $L_\text{Vainstein} $ & $L_\text{Mode-Match} $ & $L_\text{Fwd-Sctr}$ & $L_\text{Fwd-Sctr}$ \\
 \hline
0~mm & 14.6~\% & 17.1~\% & 14.1~\% &  14.8~\% \\
1~mm &  & 11.7~\% & 13.8~\% &  14.4~\% \\
2~mm &  & 11.6~\% & 13.6~\% &  14.3~\% \\
3~mm &  & 11.3~\% & 13.6~\% &  14.2~\% \\
5~mm &  & 11.2~\% & 13.4~\% &  14.1~\% \\
10~mm &  & 10.9~\% & 13.0~\% &  13.7~\% \\
25~mm &  & 10.3~\% & 12.2~\% &  12.8~\% \\
\end{tabular}
\end{ruledtabular}
\end{table}

Table~\ref{tab:results} summarized the predicted behaviour of diffraction loss as we vary the thickness $\delta$ of the thin screens over a selection of indicative values that were originally investigated in \cite{NajiTHz1}. The table compares the results obtain using the present (forward-scatter) model with those predicted by the Vainstein-based model (for $\delta=0$) \cite{DESY} and the mode-matching method (for any $\delta$) \cite{NajiTHz1}. The comparison also shows the performance of the matched Bessel mode [eigenmode $J_{0}(2.4r/a)$] versus the nearly matched Gaussian mode [$\text{exp}(-r^{2}/(0.65a)^{2}$]. 

\begin{figure}
	\includegraphics[width=0.95\columnwidth]{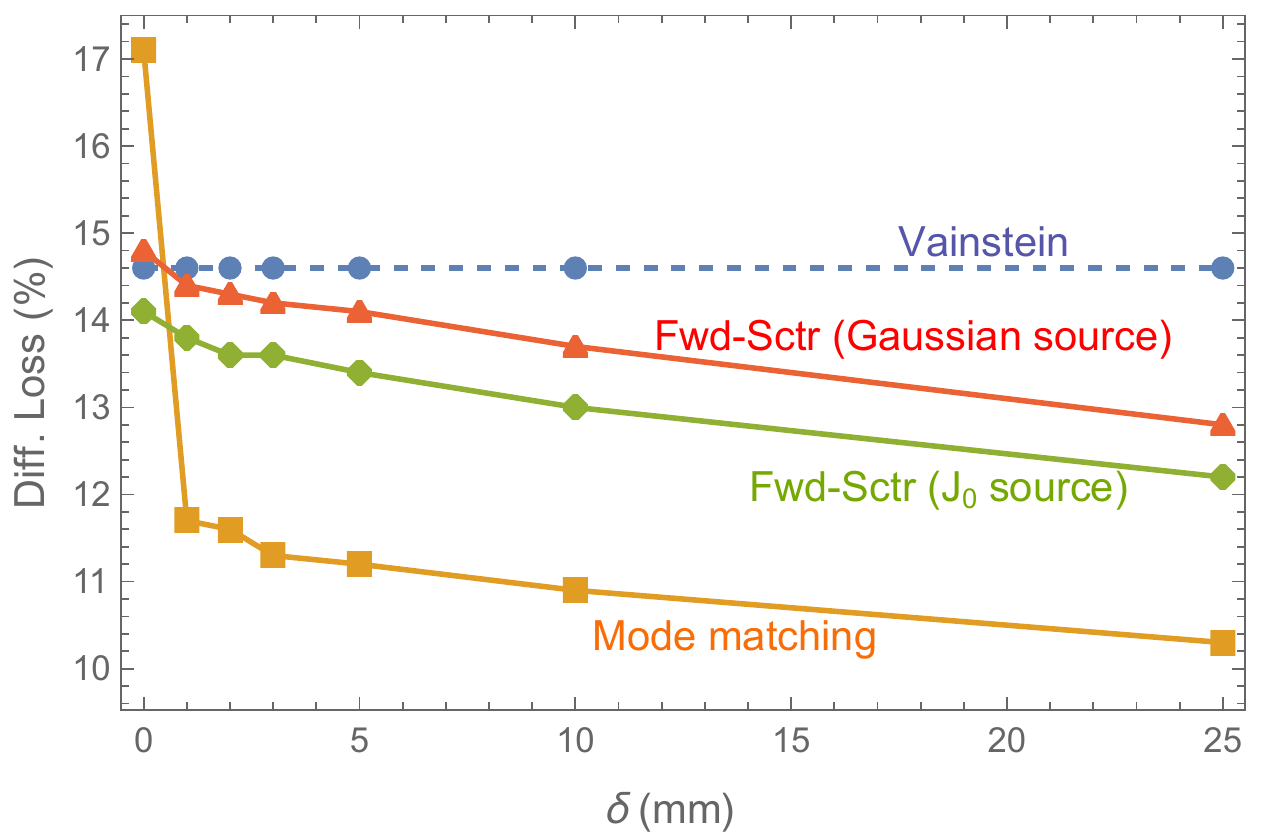}
	\caption{Plot of the data shown in Table~\ref{tab:results}.}
	\label{fig:comparisonCurves}
\end{figure}

Fig.~\ref{fig:comparisonCurves} plots the data shown in Table~\ref{tab:results}. It is clear that the diffraction loss will decrease as we gradually thicken the screens (closing the gaps between the irises), which is an expected result and in agreement with \cite{NajiTHz1}. The diffraction loss behavior shows good correlation with the trends reported by the mode-matching method. A small offset of approximately $2$\% is observed between the numerical values from the two methods. The present method seems to be in closer agreement with Vainstein's model predictions at the limit of infinitely-thin screens ($\delta\rightarrow 0$).  We note that for a practical iris line with relatively-thin screens (say, $\delta/b \leq 1\%$) the loss due to diffraction remains dominant compared to ohmic loss at the screen edges (e.g.~1\% compared to 14\%, for the example with $\delta=2$~mm, $b=33.3$~cm and $a=5.5$~cm).

\subsection{\label{sec:OtherConsiderations} Other implementation considerations}

For completeness, this section summarized how to decompose an arbitrary source signal into modal expansions in region 1 (input wavguide) and how ohmic loss can be estimated using the presented model.

\subsubsection{Decomposition of a dipole source of an arbitrary profile}

For practical implementation, we would like to represent an incident wave from a source of an arbitrary transverse profile in terms of the paraxial TE-TM modes in region 1. Since we are mainly concerned with incident waves of linear polarization (e.g.~along $\hat{x}$) and axisymmetric intensity profiles, we can explicitly write the transverse electric field of the incident wave in rectangular coordinates as
\begin{eqnarray}
      \Vec{E}&=&\hat{x}f(r)+\hat{y}0,
\end{eqnarray}
where $f(r)$ describes the profile of the source wave. 

We can now convert the field components into polar components using the transformation: $\hat{x}=\hat{r}\cos\phi-\hat{\phi}\sin\phi$ and $\hat{y}=\hat{r}\sin\phi+\hat{\phi}\cos\phi$, to write in general form
\begin{eqnarray}
\Vec{E}&=&\hat{r}f_{r}(r)\cos\phi+\hat{\phi}f_{\phi}(r)\sin\phi,
\end{eqnarray}
where here $f_{r}\equiv f$ and $f_{\phi}\equiv -f$. We can expand this field in terms of the input waveguide components (including TE and TM modes) as
\begin{equation}
\hat{r}f(r)\cos\phi-\hat{\phi}f(r)\sin\phi=\Vec{E}=\hat{r}E_{1r}+\hat{\phi}E_{1\phi} \label{BasicInputExpansion}
\end{equation}

From equations \eqref{ArhoTE}, \eqref{AphiTE}, \eqref{ArhoTM} and \eqref{AphiTM} we see that $E_{1r}$ has a $\cos\phi$ dependence, whereas $E_{1\phi}$ has a $\sin\phi$ dependence. Therefore the expansion \eqref{BasicInputExpansion} becomes
\begin{eqnarray}
-f(r)&=&\sum^{\infty}_{n=1}\frac{-A_{n}a}{r\nu'_{1n}}J_{1}(\nu'_{1n}r/a){+}B_{n}J'_{1}(\nu_{1n}r/a)\label{arb1}\\
-f(r)&=&\sum^{\infty}_{n=1}-A_{n}J'_{1}(\nu'_{1n}r/a){+}\frac{B_{n}a}{r\nu_{1n}}J_{1}(\nu_{1n}r/a)\label{arb2}
\end{eqnarray}

To find $A_{n}$: we multiply \eqref{arb1} by $J_{1}(\nu'_{1m}r/a)a/\nu'_{1m}$ and integrate with $\int^{a}_{0}r dr$; multiply \eqref{arb2} by $rJ'_{1}(\nu'_{1m}r/a)$, integrate with $\int^{a}_{0}r dr$ and take integration by parts on the second term on rhs; then add the results from the two resulting equations. This gives
\begin{eqnarray}
&&\int^{a}_{0}\left[ -f(r)rJ'_{1}(\nu'_{1m}r/a)-\frac{a}{\nu'_{1m}}f(r)J_{1}(\nu'_{1m}r/a)\right]dr \nonumber\\
&&= -\sum_{n}A_{n}\frac{a}{\nu'_{1m}}\int^{a}_{0}\left[ \frac{a}{r\nu'_{1n}}J_{1}(\nu'_{1n}r/a)J_{1}(\nu'_{1m}r/a)\right. \nonumber\\
&& \left. +rJ'_{1}(\nu'_{1n}r/a)J'_{1}(\nu'_{1m}r/a) \right]dr \label{eq:tempStar}
\end{eqnarray} 

The integral on the rhs is the same as the integral $L_{4}$ in \eqref{L4_first}, but with the replacement $r_{0}\rightarrow a$. This simplifies equation \eqref{eq:tempStar} to yield, after some algebraic manipulation, the general $A_{n}$ coefficient sought
\begin{equation}
    A_{m}=\frac{2\int^{a}_{0}rf(r) J_{0}\left(\nu'_{1m}r/a\right)dr}{a^{2}\left(1-\frac{1}{\nu'^{2}_{1m}}\right)J^{2}_{1}(\nu'_{1m})}.
\end{equation}

To find $B_{n}$: we multiply \eqref{arb1} by $rJ'_{1}(\nu_{1m}r/a)$ and integrate with $\int^{a}_{0} r dr$; multiply \eqref{arb2} by $J_{1}(\nu_{1m}r/a)a/\nu_{1m}$, integrate with $\int^{a}_{0} r dr$ and take integration by parts on the second term on rhs; then add the results from the two resulting equations. This gives
\begin{eqnarray}
&&-\int^{a}_{0}\left[ rf(r)J'_{1}(\nu_{1m}r/a)+\frac{a}{\nu_{1m}}f(r)J_{1}(\nu_{1m}r/a) \right] dr \nonumber\\
&&=\sum_{n}B_{n}\int^{a}_{0}\left[ r J'_{1}(\nu_{1n}r/a)J'_{1}(\nu_{1m}r/a) \right.\nonumber\\
&&\left. +\frac{a^{2}}{r\nu_{1n}\nu_{1m}}J_{1}(\nu_{1n}r/a)J_{1}(\nu_{1m}r/a) \right]dr
\end{eqnarray}

The integral on the rhs is nothing but $I_{1}$ in \eqref{I1_first}, if we replace $r_{0}\rightarrow a$. This leads, after simplification, to the sought $B_{n}$ coefficients as
\begin{equation}
    B_{m}=\frac{-2 \int^{a}_{0} rf(r)J_{0}\left(\nu_{1m}r/a\right)dr}{a^{2}J^{2}_{0}(\nu_{1m})},  
\end{equation}
which completes the decomposition.

\subsubsection{Carried power and attenuation constant $\alpha$ for each mode}

Each inputted mode will carry complex power in the $z$ direction, which can be derived from the following integral of Poynting's vector
\begin{eqnarray}
P_{\text{in}}&=\frac{1}{2}&\int^{a}_{0}\int^{2\pi}_{0}\Vec{E}\times\Vec{H}\cdot\hat{z}rdrd\phi\nonumber\\
&&=\frac{1}{2}\int^{a}_{0}\int^{2\pi}_{0}(E_{r}H^{*}_{\phi}-E_{\phi }H^{*}_{r})rdrd\phi.
\end{eqnarray}

Upon substitution of the field components from \eqref{ArhoTE}--\eqref{HzTE} and \eqref{ArhoTM}--\eqref{HzTM}, using integration by parts and simplifying, we find the following power in the TE or TM $l$-th mode:
\begin{eqnarray}
    P_{\text{in},l,\text{TE}}&=&\frac{{-}\pi a^{2}(\nu'^{2}_{1l}{-}1)}{4kZ_{0}}|A_{l}|^{2} \left(\frac{\nu'^{2}_{1l}}{2ka^{2}}{-}k\right)J^{2}_{0}(\nu'_{1l}) \label{TEmodePower}\\
    P_{\text{in},l,\text{TM}}&=&\frac{\pi a^{2}}{4kZ_{0}}|B_{l}|^{2}\left(\frac{\nu^{2}_{1l}}{2ka^{2}}+k\right) J^{2}_{0}(\nu_{1l})\label{TMmodePower}
\end{eqnarray}
which are real values.

Given the thin skin layer on the screen conductor at THz frequencies, we can employ the conventional perturbative method (e.g.~see \cite{pozar,Collin,slaterbook,jackson}) of finding the ohmic attenuation constant, denoted $\alpha$, from our knowledge of the tangent magnetic field at the screen-edge surface when using a perfect conductor. We can then estimate $\alpha$ as $\alpha=\frac{P_{0}}{2P_{\text{in}}}$, with $P_{0}$ being the power lost on the walls per unit length, 
\begin{equation}
    P_{0}=\frac{R_{s}}{2}\int^{2\pi}_{0}\int^{1}_{0}\left[|H_{\phi}(a,\phi,z)|^{2}+|H_{z}(a,\phi,z)|^{2}\right]a d\phi dz, \label{Po_general}
\end{equation}
where $R_{s}=\sqrt{kZ_{0}/2\sigma}$ is the surface resistance and $\sigma$ is the conductivity. Substituting in \eqref{Po_general} from \eqref{HphiTE} and \eqref{HzTE} (for the TE case), and from \eqref{HphiTM} and \eqref{HzTM}  (for the TM case), we obtain
\begin{eqnarray}
    P_{0,\text{TE}}&=&|A_{l}|^{2}\frac{aR_{s}}{2}\frac{\pi}{k^{2}Z_{0}^{2}}\left[ \frac{\nu'^{2}_{1l}}{a^{2}}J^{2}_{1}({\nu'_{1l}}) \right.\nonumber\\
    && \left. + \frac{(\nu'^{2}_{1l}/2ka^{2}-k)^{2}}{\nu'^{2}_{1l}}J^{2}_{1}(\nu'_{1l})  \right], \label{P0TE}
\end{eqnarray}
\begin{equation}
    P_{0,\text{TM}}=|B_{l}|^{2}\frac{aR_{s}}{2}\frac{\pi}{k^{2}Z_{0}^{2}} (k+\nu^{2}_{1l}/2ka^{2})^{2}J'^{2}_{1}(\nu_{1l}). \label{P0TM}
\end{equation}

We can now find $\alpha_{\text{TE},l}$ using \eqref{TEmodePower} and \eqref{P0TE}, and $\alpha_{\text{TM},l}$ using \eqref{TMmodePower} and \eqref{P0TM}, as
\begin{equation}
    \alpha_{\text{TE},l}=-\frac{R_{s}}{akZ_{0}}\frac{\nu'^{4}_{1l}/a^{2}+(\nu'^{2}_{1l}/2ka^{2}-k)^{2}}{(\nu'^{2}_{1l}-1)(\nu'^{2}_{1l}/2ka^{2}-k)}, \label{alphaTE}
\end{equation}
\begin{equation}
    \alpha_{\text{TM},l}=\frac{R_{s}}{akZ_{0}}(k+\nu^{2}_{1l}/2ka^{2}), \label{alphaTM}
\end{equation}
where we have used the fact that $J'_{1}(\nu_{1l})=[J_{0}(\nu_{1l})-J_{2}(\nu_{1l})]/2=[J_{0}(\nu_{1l})+J_{0}(\nu_{1l})]/2=J_{0}(\nu_{1l})$.

The forward-scatter model developed in the previous sections to calculate diffraction loss, which is the dominant form of loss for thin screens, may now be extended to include ohmic losses on the screen edges of each cell, by simply replacing the phase propagation term as follows
\begin{eqnarray}
    \text{TE mode: } e^{-iz\left(\frac{\nu'^{2}_{1l}}{2ka^{2}}-k\right)} &\rightarrow&  e^{- z\alpha_{\text{TE},l}}\cdot e^{-iz\left(\frac{\nu'^{2}_{1l}}{2ka^{2}}-k\right)},\nonumber\\
    \text{TM mode: } e^{-iz\left(\frac{\nu^{2}_{1l}}{2ka^{2}}-k\right)} &\rightarrow&  e^{- z\alpha_{\text{TM},l}}\cdot e^{-iz\left(\frac{\nu'^{2}_{1l}}{2ka^{2}}-k\right)}.\nonumber
\end{eqnarray}

\section{\label{sec:conclusions} Conclusions}
Field and transient analysis for the paraxial propagation of a short THz pulse in an oversized iris line has been presented, where we have exploited the shortness of the pulse and the largeness of the structure to approximate the problem as a forward-scatter problem. The developed model allows for the study of transients along the iris line and for the investigations of propagation performance for a finite length of line and an arbitrary driving source. This is contrasted to the steady-state eigensolutions found by previous methods, such as the mode-matching method. The developed model boosts the computational speed by an order of magnitude compared to the mode-matching method and offers a convenient computational platform that can be easily extended to examine sensitivity to mechanical tolerances or geometrical defects. 

\begin{acknowledgments}
The authors would like to thank Robert Warnock, Karl Bane, Emilio Nanni and Zhirong Huang, of SLAC, Stanford University, for several interesting discussions in relation to this research. This work was supported
by the US Department of Energy (contract number DE-AC02-76SF00515).

\end{acknowledgments}

\appendix

\section{\label{Appx1} Wave equation solution derivations}

Solving Maxwell's curl equations together in free space gives the vector wave equation for the electric field $\mathcal{E}$
\begin{equation}
    \nabla^{2}\Vec{\mathcal{E}}-\frac{1}{c^2}\frac{\partial^{2}}{\partial t^{2}}\Vec{\mathcal{E}}=0. \label{VecWaveEq1}
\end{equation}

Given the time-harmonic dependence $e^{-i\omega t}$, we may write the field in phasor form as $\Vec{\mathcal{E}}(r,z,t)=\text{Re}\left[ \Vec{E}(r,\phi,z) e^{-i\omega t} \right]$, converting (\ref{VecWaveEq1}) into the vector Helmholtz equation
\begin{equation}    
\nabla^{2}\Vec{E}+k^{2}\Vec{E}=0. \label{Helmholtz_E_1}
\end{equation}

We assume the field $\Vec{E}=\hat{r}E_{r}+\hat{\phi}E_{\phi}+\hat{z}E_{z}$ to have the form
\begin{eqnarray}
    E_{ r}&=&A_{ r}( r,\phi,z)e^{ikz},\\
    E_{\phi}&=&A_{\phi}( r,\phi,z)e^{ikz},\\
    E_{z}&=&A_{z}( r,\phi,z)e^{ikz},
\end{eqnarray}
where the functions $A_{r},A_{\phi}$ and $A_{z}$ represent the complex field envelopes. 

Let us expand the vector Laplacian into its orthogonal components and solve for each component in the Helhmotlz equation (\ref{Helmholtz_E_1}) separately.  Starting with the $ r$ component, (\ref{Helmholtz_E_1})  reduces to\begin{equation} \label{tempEq_1}
    \underbrace{\frac{1}{ r}\frac{\partial}{\partial r} r\frac{\partial E_{ r}}{\partial  r}}_{\text{term 1}}+\underbrace{\frac{1}{ r^{2}}\frac{\partial^{2}E_{ r}}{\partial\phi^{2}}}_{\text{term 2}}+\underbrace{\frac{\partial^{2}E_{ r}}{\partial z^{2}}}_{\text{term 3}}-\frac{E_{ r}}{ r^{2}}-\frac{2}{ r^{2}}\frac{\partial E_{\phi}}{\partial \phi}+k^{2}E_{ r}=0
\end{equation}

Since we know that the azimuthal dependence is dipolar (i.e.~either of the form $\sin\phi$ or $\cos\phi$), we know that the operator $\partial^{2}/\partial\phi^{2}$ is equivalent to multiplication by scalar $-1$, and we can expand the terms 1--3 in (\ref{tempEq_1}) to arrive at
\begin{eqnarray}
\text{term 1 }&=&e^{ikz}\left( \frac{1}{ r} \frac{\partial A_{ r}}{\partial r} + \frac{\partial^{2} A_{ r}}{\partial r^{2}}\right)\\
\text{term 2 }&=&-e^{ikz}\frac{A_{ r}}{ r^{2}}\\
\text{term 3 }&\approx&e^{ikz}\left( 2ik \frac{\partial A_{ r}}{\partial z} - k^{2}A_{ r} \right)
\end{eqnarray}
where for term 3 we have also assumed a slowly-varying envelope for $A_{r}$ along $z$ (implying $\frac{\partial^{2}A_{ r}}{\partial z^{2}}\ll k^{2}A_{ r}$). Substituting these terms back into (\ref{tempEq_1}) and assuming the following separable form for the field envelopes (justifiable by the separability of the Helmholt equation itself in cylindrical coordinates \cite{Morse} as well as the axisymmetrical geometry of the iris line)
\begin{eqnarray}
A_{ r}( r,\phi,z)=R( r)Z(z)\cos\phi,\\
A_{\phi}( r,\phi,z)=\hat{R}( r)\hat{Z}(z)\sin\phi,
\end{eqnarray}
give us, after algebraic simplification, the following equation
\begin{equation}
    ZR''+\frac{ZR'}{ r}-\frac{2ZR}{ r^{2}}+2ikRZ'=\frac{2\hat{R}\hat{Z}}{ r^{2}}. \label{ZR1}
\end{equation}

Following similar steps for the $\phi$ component we obtain
\begin{equation}
    \frac{1}{ r}\frac{\partial}{\partial r} r\frac{\partial E_{\phi}}{\partial  r}+\frac{1}{ r^{2}}\frac{\partial^{2}E_{\phi}}{\partial\phi^{2}}+\frac{\partial^{2}E_{\phi}}{\partial z^{2}}-\frac{E_{\phi}}{ r^{2}}+\frac{2}{ r^{2}}\frac{\partial E_{ r}}{\partial \phi}+k^{2}E_{\phi}=0,
\end{equation}
which, upon assuming a slowly-varying envelope for $A_{\phi}$ envelope along $z$ (implying $\frac{\partial^{2}A_{\phi}}{\partial z^{2}}\ll k^{2}A_{\phi}$) and simplifying, yields the following equation
\begin{equation}
    \hat{Z}\hat{R}''+\frac{\hat{Z}\hat{R}'}{ r}-\frac{2\hat{Z}\hat{R}}{ r^{2}}+2ik\hat{R}\hat{Z}'=\frac{2RZ}{ r^{2}}. \label{ZR2}
\end{equation}

We notice the symmetry in form between equations (\ref{ZR1}) and (\ref{ZR2}). Given that the envelopes $A_{r},A_{\phi}$ are both slowly varying in $z$ and given the axisymmetrical nature of the structure, we expect the variation and derivatives in $z$ to be uncoupled of those in $r$ and $\phi$, allowing us to make the assumption that $Z=\hat{Z}$. We proceed to take the sum and the difference of equations (\ref{ZR1}) and (\ref{ZR2}), to yield the equation pair
\begin{eqnarray}
Z(\hat{R}{-}R)''{+}\frac{Z}{ r}(\hat{R}{-}R)'{+}2ikZ'(\hat{R}{-}R)&=&0,\label{diff}\\
Z(\hat{R}{+}R)''{+}\frac{Z}{ r}(\hat{R}{+}R)'{+}2ikZ'(\hat{R}{+}R)&=&\frac{4Z}{ r^{2}}(\hat{R}{+}R).\label{sum}
\end{eqnarray}

Let us derive the solutions for (\ref{sum}) and (\ref{diff}) of Section~\ref{sec:BasicFields}. If we define the function $f( r)\equiv R-\hat{R}$, then (\ref{diff}) can be reduced to a separable equation, with the lhs in $ r$ only, while the rhs in $z$ only, as
\begin{equation}
    \frac{f''}{f}+\frac{1}{ r}\frac{f'}{f}=-2ik\frac{Z'}{Z}=-c_{o}^{2},
\end{equation}
where $c_{0}$ is an arbitrary separation constant to be determined later by the boundary conditions. We recognize the lhs as a Bessel equation in $ r$, while the rhs as a harmonic equation in $z$,
\begin{eqnarray}
f''+\frac{1}{ r}f'+c_{0}^{2}f=0 &\Rightarrow& f( r)=AJ_{0}(c_{0} r),\\
\frac{Z'}{Z}=-\frac{2c_{0}^{2}}{2k} &\Rightarrow& Z(z)=e^{-iz\frac{c^{2}_{0}}{2k}},
\end{eqnarray}
where $A$ is a constant.

Similarly, let us define the function $g( r)\equiv R+\hat{R}$, so that (\ref{sum}) can be reduced to the separable equation
\begin{eqnarray}
\frac{g''}{g}+\frac{1}{ r}\frac{g'}{g}-\frac{4}{ r^{2}}=-2ik\frac{Z'}{Z}=-d_{0}^{2},
\end{eqnarray}
where $d_{0}$ is an arbitrary separation constant. We recognize the lhs as a Bessel equation in $ r$, while the rhs as a harmonic equation in $z$,
\begin{eqnarray}
g''+\frac{1}{ r}g'+\left( d_{0}^{2} -\frac{2^{2}}{ r^{2}}\right) g=0 &\Rightarrow& g( r)=A' J_{2}(d_{0} r),\\
\frac{Z'}{Z}=-\frac{2d_{0}^{2}}{2k} &\Rightarrow& Z(z)=e^{-iz\frac{d^{2}_{0}}{2k}},
\end{eqnarray}
where $A'$ is a constant.

We put $c_{0}=d_{0}$ to have the $z$ dependences equal, giving us in summary
\begin{eqnarray}
 Z(z)&=&e^{-iz\frac{d^{2}_{0}}{2k}},\\
 R+\hat{R}&=&g( r)=A'J_{2}(c_{0} r),\\
 R-\hat{R}&=&f( r)=AJ_{0}(c_{0} r).
\end{eqnarray}

Given that $2R=f+g$, that $J_{1}(z)/z=J_{0}(z)+J_{2}(z)$, and that we can write the constant $A'$ as $A+B$, where $B$ is some constant, we can simplify and obtain
\begin{equation}
    R( r)=\frac{A}{c_{0} r}J_{1}(c_{0} r)+BJ_{2}(c_{0} r).
\end{equation}

Similarly, given that $2\hat{R}=g-f$, that $2J'_{1}(z)=J_{0}(z)-J_{2}(z)$, and that we can write the constant $A'$ as $A+B$, we can simplify and obtain
\begin{equation}
    \hat{R}( r)=-A J'_{1}(c_{0} r)+BJ_{2}(c_{0} r).
\end{equation}

It is interesting to note in passing that the functions $R$ and $\hat{R}$ have different forms in general, even though equations (\ref{ZR1}) and (\ref{ZR2}) exhibited symmetry.  We can now collect the different terms and substitute in the definitions of the envelopes $A_{r}, A_{\phi}$ to arrive at 
give the field envelopes $A_{ r}, A_{\phi}$ and $A_{z}$ as
\begin{eqnarray}
   A_{ r}( r,\phi,z)&=&\left[ \frac{A}{c_{0} r} J_{1}(c_{0} r){+}B J_{2}(c_{0} r) \right]\cos\phi e^{-iz\frac{c^{2}_{0}}{2k}},\label{Arho}\\
   A_{\phi}( r,\phi,z)&=&\left[ -A J'_{1}(c_{0} r){+}B J_{2}(c_{0} r) \right]\sin\phi e^{-iz\frac{c^{2}_{0}}{2k}},\label{Aphi}\\
   A_{z}( r,\phi,z)&=&\frac{iBc_{0}}{k} J_{1}(c_{0} r)\cos\phi e^{-iz\frac{c^{2}_{0}}{2k}}, \label{Az}
\end{eqnarray}
where $A, B$ are arbitrary complex constants and to find $A_{z}$ we used Maxwell's divergence equation $(\nabla\cdot\Vec{E}=0)$, with the assumption of slowly-varying envelope $A_{z}$ along $z$ (implying $\frac{\partial A_{z}}{\partial z}\ll k A_{z}$), as shown below. 

We have
\begin{equation}
    A_{z}=\frac{i}{ r k}\left( A_{ r}+ r \frac{\partial A_{ r}}{\partial r}+\frac{\partial A_{\phi}}{\partial\phi} \right),
\end{equation}

and the terms $ r \frac{\partial A_{ r}}{\partial r}$ and $\frac{\partial A_{\phi}}{\partial\phi}$ can be shown to reduce to (in shorthand notation)
\begin{eqnarray}
     r \frac{\partial A_{ r}}{\partial r}&=&\left( AJ'_{1}-\frac{A}{c_{0} r}J_{1}+B c_{0} r J'_{2} \right)\cos\phi e^{-iz\frac{c^{2}_{0}}{2k}},\\
    \frac{\partial A_{\phi}}{\partial\phi}&=&\left( -AJ'_{1}+BJ'_{2} \right)\cos\phi e^{-iz\frac{c^{2}_{0}}{2k}},
\end{eqnarray}
leading us to the final $A_{z}$ expression
\begin{equation}
    A_{z}=\frac{iBc_{0}}{k} J_{1}(c_{0} r)\cos\phi e^{-iz\frac{c^{2}_{0}}{2k}}, \label{Az__}
\end{equation}
which was cited earlier.

Note that the magnetic field can be derived from the electric field found above via Maxwell's curl equation $\Vec{H}=-\frac{i}{\omega\mu}\nabla\times\Vec{E}$. Upon writing the curl components explicitly, we obtain (after simplifying)
\begin{eqnarray}
    H_{ r}&=&\frac{-i}{\omega\mu}\left( \frac{1}{ r} \frac{\partial A_{z}}{\partial\phi}-\frac{\partial A_{\phi}}{\partial z}-ikA_{\phi}\right)e^{+ikz},\label{H_rho_general}\\
    H_{\phi}&=&\frac{-i}{\omega\mu}\left(\frac{\partial A_{ r}}{\partial z} +ikA_{ r}-\frac{\partial A_{z}}{\partial  r}\right)e^{+ikz},\label{H_phi_general}\\
    H_{z}&=&\frac{-i}{\omega\mu}\frac{1}{ r}\left( A_{\phi}+ r \frac{\partial A_{\phi}}{\partial  r}-\frac{\partial A_{ r}}{\partial\phi} \right)e^{+ikz},\label{H_z_general}
\end{eqnarray}
which are the general components of the magnetic field. These components will be further specified for the TE and TM modes below.

Let us now impose the boundary conditions $A_{z}=0$ and $A_{\phi}=0$ at $ r=q$ (for waveguide sections) and at $ r=r_{0}$ (for the cavity section).  Starting at the waveguide wall case ($ r=a$), from (\ref{Az}) we see that it will require one of two nontrivial solutions: either $B=0$, giving $A_{z}=0$, which corresponds to the TE case; or $c_{0}=\nu_{1n}/a$, with $B\neq 0$, which corresponds to the TM case, where $\nu_{1n}$ is the $n$th root of the $J_{1}$ function. 

We can now proceed with applying the boundary condition also on equation (\ref{Aphi}) for each of the TE and TM cases as follows. 

\textbf{TE case:} here $B=0, A_{z}=0$ generally and we still need to determine $c_{0}$ by requiring equation (\ref{Aphi}) to be zero at $r=a$. This leads to either requiring $A=0$ or $c_{0}=\nu'_{1n}/a$. We reject the former option because it leads to all fields vanishing (trivial solution) and adopt the latter one. Substituting back into our fields and simplifying, we find the final TE fields given in equation (\ref{ArhoTE})--(\ref{HzTE}). For the cavity section, replace $a$ by $r_{0}$, which results in the expansion equations (\ref{ArhoTEc})--(\ref{HzTEc}). Note that $Z_{0}=\sqrt{\mu_{0}/\epsilon_{0}}$ denotes the free-space impedance, and the magnetic field components are found from the electric field components using (\ref{H_rho_general})--(\ref{H_z_general}).

\textbf{TM case:} here $c_{0}=\nu'_{1n}/a$, with $B\neq 0, A_{z}\neq 0$ generally, and for equation (\ref{Aphi}) to be zero at $ r=a$ we need
\begin{equation}
    BJ_{2}(\nu_{1n})-AJ'_{1}(\nu_{1n})=0. \label{cond}
\end{equation}
We have $J'_{1}(\nu_{1n})=J_{0}(\nu_{1n})-\frac{1}{\nu_{1n}a}J_{1}(\nu_{1n})=J_{0}(\nu_{1n})-0=J_{0}(\nu_{1n})$ here, however, so (\ref{cond}) becomes
\begin{equation}
    BJ_{2}(\nu_{1n})-AJ_{0}(\nu_{1n})=0. 
\end{equation}
Since $\nu_{1n}$ are the zeros of the $J_{1}$ functions, and since we know that the zeros of the functions $J_{0}, J_{1}$ and $J_{2}$ are non-overlapping \cite{NIST}, we know that the only solution for this equation is obtainable by finding the suitable values for $A,B$ constants. This gives us $A=B X_{a}$, where $X_{a}=J_{2}(c_{0}a)/J_{0}(c_{0}a)$. However, we know that $J_{2}=J_{0}-2J'_{1}$ and $J'_{1}(\nu_{1n})=J_{0}(\nu_{1n})$, since $J_{1}(\nu_{1n})=0$, which leads to $X_{a}=-1 \Rightarrow A=-B$. Substituting back into our fields and simplifying, we find the final TM fields given in equation (\ref{ArhoTM})--(\ref{HzTM}). For the cavity section, replace $a$ by $r_{0}$, which results in the expansion equations (\ref{ArhoTMc})--(\ref{HzTMc}).

\section{\label{Appx2} Analytical calculation of integrals}
Integral $I_{1}$ from \eqref{I1_first} can be calculated as follows. Simplify the integral
\begin{eqnarray} \label{I1}
I_{1}&=&\int_0^{r_{0}} r\,dr J'_{1}\left( \nu_{1n}\frac{r}{r_{0}}\right)
  J'_{1}\left(\nu_{1m}\frac{r}{r_{0}}\right) \nonumber\\
  &&+
  \frac{r_{0}^2}{\nu_{1n}\nu_{1m}}
  \int_0^{r_{0}} \frac{dr}{r}
  J_{1}\left( \nu_{1n}\frac{r}{r_{0}}\right)
  J_{1}\left( \nu_{1m}\frac{r}{r_{0}}\right),
\end{eqnarray}
by expanding each term according to the identity $2J'_{1}\left( \nu_{1n}\frac{r}{r_{0}}\right)=J_{0}\left( \nu_{1n}\frac{r}{r_{0}}\right)-J_{2}\left( \nu_{1n}\frac{r}{r_{0}}\right)$ or the identity $(2r_{0}/\nu_{1n}r)J_{1}\left( \nu_{1n}\frac{r}{r_{0}}\right)=J_{0}\left( \nu_{1n}\frac{r}{r_{0}}\right)+J_{2}\left( \nu_{1n}\frac{r}{r_{0}}\right)$, then grouping the terms to give
\begin{eqnarray}
    I_{1}&=&\frac{1}{2}\int^{r_{0}}_{0}r J_{0}\left( \nu_{1n}\frac{r}{r_{0}}\right) J_{0}\left( \nu_{1m}\frac{r}{r_{0}}\right) \nonumber\\
    &&+\frac{1}{2}\int^{r_{0}}_{0}r J_{2}\left( \nu_{1n}\frac{r}{r_{0}}\right) J_{2}\left( \nu_{1m}\frac{r}{r_{0}}\right)\\
    &=&\begin{cases}
    0, & m\neq n\\
    \frac{r_{0}^{2}}{2}J^{2}_{0}(\nu_{1m})=\frac{r_{0}^{2}}{2}J'^{2}_{1}(\nu_{1m}), & m=n
    \end{cases}\nonumber\\
    &=&\delta_{mn}\frac{r_{0}^{2}}{2}J^{2}_{0}(\nu_{1m})=\delta_{mn}\frac{r_{0}^{2}}{2}J'^{2}_{1}(\nu_{1m}), \label{I1_result}
\end{eqnarray}
    since we have $J_{0}(\nu_{1n})=J'_{1}(\nu_{1n})$. Here, $\delta_{mn}$ denotes Kronecker's delta function.

Integral $I_{2}$ from \eqref{I2_first} can be calculated as follows. First, let us simplify the integral 
\begin{eqnarray}\label{I2}
        I_{2}&=&\frac{r_{0}}{\nu'_{1m}} \left[ \int_0^{r_{0}}   \frac{r_{0}}{\nu'_{1n}{r}} dr
  J_{1}\left(\nu'_{1n}\frac{r}{r_{0}}\right) 
  J_{1}\left(\nu'_{1m}\frac{r}{r_{0}}\right) \right.\nonumber\\
  &&  \left. +
  \int_0^{r_{0}}  dr \frac{\nu'_{1m}r}{r_{0}}
  J'_{1}\left(\nu'_{1n}\frac{r}{r_{0}}\right)
  J'_{1}\left( \nu'_{1m}\frac{r}{r_{0}}\right) \right],
    \end{eqnarray}
by using the shorthand notation $J_{1}(\nu'_{1n}r/r_{0})\rightarrow J_{1n}, J_{1}(\nu'_{1m}r/r_{0})\rightarrow J_{1m}$, and so forth. Then, using the Bessel identities of the form $(r\nu'_{1m}/r_{0})J'_{1m}=(\nu'_{1m}r/r_{0})J_{0m}-J_{1m}$ and $(r_{0}/r\nu_{1n})J_{1n}=J_{0n}-J'_{1n}$ a few times, and rearranging the terms, we can finally arrive at
    \begin{eqnarray}
        I_{2}&=&\frac{r_{0}}{\nu'_{1m}} \int_0^{r_{0}} dr\left[J_{0n}J_{1m}-J'_{1n}\left( 2 J_{1m}-\frac{\nu'_{1m}r}{r_{0}}J_{0m} \right)\right]\nonumber\\
        &=&\frac{r_{0}}{\nu'_{1m}} \int_0^{r_{0}} dr\frac{\nu'_{1m}r}{r_{0}} \left[\frac{1}{2}J_{0n}J_{0m}+ \frac{1}{2}J_{0n}J_{2m}-J'_{1n}J_{2m}\right]\nonumber\\
        &=&\frac{r_{0}}{\nu'_{1m}} \int_0^{r_{0}} dr\frac{\nu'_{1m}r}{2r_{0}} \left[J_{0n}J_{0m}+J_{2n}J_{2m}\right]\nonumber\\
        &=&\frac{r_{0}}{\nu'_{1m}}\begin{cases}
    0, & m\neq n\\
    \frac{r_{0}\nu'_{1m}}{2}J^{2}_{0}(\nu'_{1m})(\nu'^{2}_{1m}-1), & m=n
    \end{cases}\nonumber\\
    &=&\delta_{mn}\frac{r_{0}\nu'_{1m}}{2}(\nu'^{2}_{1m}-1)J^{2}_{0}(\nu'_{1m})\nonumber\\
    &=&\delta_{mn}\frac{r_{0}\nu'_{1m}}{2}(1-1/\nu_{1m}^{2})J^{2}_{1}(\nu'_{1m}). \label{I2_result}
    \end{eqnarray}

Integral $I_{3}$ from \eqref{I3_first} can be calculated with the help of Bessel identities of the form $(r\nu_{1n}/r_{0})J'_{1}(r\nu_{1n}/r_{0})=(\nu_{1n}r/r_{0})J_{0}(r\nu_{1n}/r_{0})-J_{1}(r\nu_{1n}/r_{0})$ and regrouping the terms, to yield
    \begin{eqnarray}
I_{3}&=&\int_0^a dr\left[r
  J'_{1}\left( \nu_{1l}\frac{r}{a}\right)
  J'_{1}(\nu_{1m}\frac{r}{r_{0}}) \right. \nonumber\\
  && \left. +  r\frac{a}{\nu_{1l}r}
  J_{1}\left( \nu_{1l}\frac{r}{a}\right)
  \frac{r_{0}}{\nu_{1m}r} J_{1}\left( \nu_{1m}\frac{r}{r_{0}}\right)
  \right]\label{I3}\\
  &=&\int_0^a dr\left\{ r
  J'_{1}\left( \nu_{1l}\frac{r}{a}\right)
  J'_{1}(\nu_{1m}\frac{r}{r_{0}}) +  r\left[ J_{0}(\nu_{1l}r/a) \right. \right. \nonumber\\
  && \left. \left. -J'_{1}(\nu_{1l}\frac{r}{a})\right]\left[ J_{0}(\nu_{1m}\frac{r}{r_{0}})-J'_{1}(\nu_{1m}\frac{r}{r_{0}})\right]
  \right\}\nonumber,
    \end{eqnarray}
which, upon expanding the products, regrouping the factors of $J'_{1}(\nu_{1m}r/r_{0})$ and $J'_{1}(\nu_{1l}r/r_{0})$, then simplifying using the same Bessel identities used above, we obtain two simple integrands (one of which is a perfect differential). Namely,
    \begin{eqnarray}
        I_{3}&=&-\int^{a}_{0}dr\left\{ \frac{r_{0}a}{\nu_{1l}\nu_{1m}}\frac{d}{dr}\left[ J_{1}(\nu_{1l}\frac{r}{a})J_{1}(\nu_{1m}\frac{r}{r_{0}}) \right] \right. \nonumber\\
        && \left. -rJ_{0}(\nu_{1l}\frac{r}{a})(\nu_{1m}\frac{r}{r_{0}})  \right\}\nonumber\\
        &=&0+\int^{a}_{0}r\,dr J_{0}(\nu_{1l}\frac{r}{a})(\nu_{1m}\frac{r}{r_{0}})  \nonumber\\
        &=&\frac{a \nu_{1m} J_{0}(\nu_{1l})J_{1}(\nu_{1m}a/r_{0})}{r_{0}(\nu^{2}_{1m}/r_{0}^{2}-\nu^{2}_{1l}/a^{2})}. \label{I3_result}
    \end{eqnarray}

Using shorthand notation of the form $J_{0}(\nu'_{1n}r/r_{0})\rightarrow J_{0n'}, J_{1}(\nu_{1m}r/r_{0})\rightarrow J_{1m}$, and so forth, we can rewrite and calculate integrals $L_{3}$, $L_{4}$, $\hat{L}_{3}$ and $\hat{L}_{4}$ as follows.
\begin{widetext}
\begin{eqnarray}
    L_{3}&=&\frac{\nu'_{1m}}{r_{0}}\left[ \int^{a}_{0}rJ'_{1}(\nu'_{1l}\frac{r}{a})J'_{1}(\nu'_{1m}\frac{r}{r_{0}})dr  +\int^{a}_{0}r\frac{a}{\nu'_{1l}r}J_{1}(\nu'_{1l}\frac{r}{a})\frac{r_{0}}{\nu'_{1m}r}J_{1}(\nu'_{1m}\frac{r}{r_{0}})dr  \right]\nonumber\\
    &=&\frac{\nu'_{1m}}{r_{0}}\left[ \int^{a}_{0}r\frac{1}{2}(J_{0l'}-J_{2l'})\frac{1}{2}(J_{0m'}-J_{2m'})dr  + \int^{a}_{0}r\frac{1}{2}(J_{0l'}+J_{2l'})\frac{1}{2}(J_{0m'}+J_{2m'})dr \right]\nonumber\\
    &=&\frac{\nu'_{1m}}{4r_{0}}\int^{a}_{0} r\left(2J_{0l'}J_{0m'}+2J_{2l'}J_{2m'} \right)dr \nonumber\\
    &=&\frac{\nu'_{1m}}{2r_{0}} \left[  \int^{a}_{0} rJ_{0}(\nu'_{1l}\frac{r}{a})J_{0}(\nu'_{1m}\frac{r}{r_{0}})dr  + \int^{a}_{0} rJ_{2}(\nu'_{1l}\frac{r}{a})J_{2}(\nu'_{1m}\frac{r}{r_{0}})dr \right],\nonumber\\
     &=&\frac{r_{0} \nu'_{1m}\nu'_{1l}J_{1}(\nu'_{1l})J'_{1}(\nu'_{1m}a/r_{0})}{(\nu'^{2}_{1l}r_{0}^{2}/a^{2}-\nu'^{2}_{1m})}. \label{L3found}
\end{eqnarray}
\begin{eqnarray}
    L_{4}&=&\int^{r_{0}}_{0}dr\frac{r_{0}}{\nu'_{1n}r}J_{1}(\nu'_{1n}\frac{r}{r_{0}})J_{1}(\nu'_{1m}\frac{r}{r_{0}})+\int^{r_{0}}_{0}dr\frac{\nu'_{1m}r}{r_{0}}J'_{1}(\nu'_{1n}\frac{r}{r_{0}})J'_{1}(\nu'_{1m}\frac{r}{r_{0}})\\
    &=&\int^{r_{0}}_{0}dr\left[ J_{0}(\nu'_{1n}\frac{r}{r_{0}})-J'_{1}(\nu'_{1n}\frac{r}{r_{0}}) \right] J_{1}(\nu'_{1m}\frac{r}{r_{0}})+\int^{r_{0}}_{0}dr\left[ \frac{
    \nu'_{1m}r}{r_{0}} J_{0}(\nu'_{1m}\frac{r}{r_{0}})-J_{1}(\nu'_{1m}\frac{r}{r_{0}}) \right]J'_{1}(\nu'_{1n}\frac{r}{r_{0}})\nonumber\\
    &=&\int^{r_{0}}_{0}dr\left(J_{0n'}J_{1m'}-J'_{1n'}J_{1m'}+\frac{\nu'_{1m}r}{r_{0}}J'_{1n'}J_{0m'}-J'_{1n'}J_{1m'}\right)\nonumber\\
    &=&\int^{r_{0}}_{0}dr\left[ J_{0n'}J_{1m'}-J'_{1n'}\left( 2 J_{1m'}-\frac{\nu'_{1m}r}{r_{0}}J_{0m'} \right)   \right]\nonumber\\
    &=&\int^{r_{0}}_{0}dr\left[ J_{0n'}J_{1m'}-\frac{\nu'_{1m}r}{r_{0}}J'_{1n'}J_{2m'}   \right] \nonumber\\
    &=&\int^{r_{0}}_{0}dr\frac{\nu'_{1m}r}{2r_{0}}\left( J_{0n'}J_{0m'}+J_{0n'}J_{2m'}-J_{0n'}J_{2m'}+J_{2n'}J_{2m'}
    \right)\\
    &=&\frac{\nu'_{1m}}{2r_{0}}\left[ \int^{r_{0}}_{0} r J_{0}(\nu'_{1n}\frac{r}{r_{0}})J_{0}(\nu'_{1m}\frac{r}{r_{0}})dr+\int^{r_{0}}_{0} r J_{2}(\nu'_{1n}\frac{r}{r_{0}})J_{2}(\nu'_{1m}\frac{r}{r_{0}})dr\right]\\
    &=&\begin{cases}
    0, & n\neq m\\
    \frac{\nu'_{1m}}{2r_{0}}\frac{1}{2}r_{0}^{2}\left[ J^{2}_{1}(\nu'_{1m})(1+1/\nu'^{2}_{1m})+J^{2}_{2}(\nu'_{1m})-J_{1}(\nu'_{1m})J_{3}(\nu'_{1m}) \right], &n=m
    \end{cases}\\
    &=&\delta_{mn}\frac{r_{0}}{2}(\nu'_{1m}-1/\nu'_{1m})J^{2}_{1}(\nu'_{1m}),\label{L4found}
\end{eqnarray}
where the last equality was reached using the fact that $J_{3}(\nu'_{1m})=4 J_{2}(\nu'_{1m})/\nu'_{1m}-J_{1}(\nu'_{1m})$ and that $J_{0}(\nu'_{1m})=J_{2}(\nu'_{1m})=J_{1}(\nu'_{1m})/\nu'_{1m}$.

\begin{eqnarray}
    \hat{L}_{3}&=&\frac{\nu'_{1m}}{a}\left[ \int^{a}_{0}rJ'_{1}(\nu'_{1l}\frac{r}{r_{0}})J'_{1}(\nu'_{1m}\frac{r}{a})dr+\int^{a}_{0}r\frac{r_{0}}{\nu'_{1l}r}J_{1}(\nu'_{1l}\frac{r}{r_{0}})\frac{a}{\nu'_{1m}r}J_{1}(\nu'_{1m}\frac{r}{a})dr  \right]\\
    &=&\frac{\nu'_{1m}}{a}\left[ \int^{a}_{0}r\frac{1}{2}(J_{0l'}-J_{2l'})\frac{1}{2}(J_{0m'}-J_{2m'})dr + \int^{a}_{0}r\frac{1}{2}(J_{0l'}+J_{2l'})\frac{1}{2}(J_{0m'}+J_{2m'})dr \right]\\
    &=&\frac{\nu'_{1m}}{4a}\int^{a}_{0} r\left(2J_{0l'}J_{0m'}+2J_{2l'}J_{2m'} \right)dr\\
     &=&\frac{\nu'_{1m}}{2a} \left[  \int^{a}_{0} rJ_{0}(\nu'_{1l}\frac{r}{a})J_{0}(\nu'_{1m}\frac{r}{a})dr + \int^{a}_{0} rJ_{2}(\nu'_{1l}\frac{r}{a})J_{2}(\nu'_{1m}\frac{r}{a})dr \right]\\
     &=&\frac{a r_{0}^{2} \nu'^{2}_{1m}J_{1}(\nu'_{1m})J'_{1}(\nu'_{1l}a/r_{0})}{a^{2}(\nu'^{2}_{1m}r_{0}^{2}/a^{2}-\nu'^{2}_{1l})}. \label{L3found2}
\end{eqnarray}
It is interesting to note here how the integral result leading to \eqref{L3found2} differs from that leading to \eqref{L3found} by the interchange of $\nu'_{1m}\leftrightarrow \nu'_{1m}$ (instead of $r_{0}\leftrightarrow a$).
\begin{eqnarray}
    \hat{L}_{4}&=&\int^{a}_{0}dr\frac{a}{\nu'_{1n}r}J_{1}(\nu'_{1n}\frac{r}{a})J_{1}(\nu'_{1m}\frac{r}{a})+\int^{a}_{0}dr\frac{\nu'_{1m}r}{a}J'_{1}(\nu'_{1n}\frac{r}{a})J'_{1}(\nu'_{1m}\frac{r}{a})\\
    &=&\int^{a}_{0}dr\left[ J_{0}(\nu'_{1n}\frac{r}{a})-J'_{1}(\nu'_{1n}\frac{r}{a}) \right] J_{1}(\nu'_{1m}\frac{r}{a})+\int^{a}_{0}dr\left[ \frac{
    \nu'_{1m}r}{a} J_{0}(\nu'_{1m}\frac{r}{a})-J_{1}(\nu'_{1m}\frac{r}{a}) \right]J'_{1}(\nu'_{1n}\frac{r}{a})\nonumber\\
    &=&\int^{a}_{0}dr\left(J_{0n'}J_{1m'}-J'_{1n'}J_{1m'}+\frac{\nu'_{1m}r}{a}J'_{1n'}J_{0m'}-J'_{1n'}J_{1m'}\right)\nonumber\\
    &=&\int^{a}_{0}dr\left[ J_{0n'}J_{1m'}-J'_{1n'}\left( 2 J_{1m'}-\frac{\nu'_{1m}r}{a}J_{0m'} \right)   \right]\nonumber\\
    &=&\int^{a}_{0}dr\left[ J_{0n'}J_{1m'}-\frac{\nu'_{1m}r}{a}J'_{1n'}J_{2m'}   \right] \nonumber\\
    &=&\int^{a}_{0}dr\frac{\nu'_{1m}r}{2a}\left( J_{0n'}J_{0m'}+J_{0n'}J_{2m'}-J_{0n'}J_{2m'}+J_{2n'}J_{2m'}
    \right)\\
    &=&\frac{\nu'_{1m}}{2a}\left[ \int^{a}_{0} r J_{0}(\nu'_{1n}\frac{r}{a})J_{0}(\nu'_{1m}\frac{r}{a})dr+\int^{a}_{0} r J_{2}(\nu'_{1n}\frac{r}{a})J_{2}(\nu'_{1m}\frac{r}{a})dr\right]\\
    &=&\begin{cases}
    0, & n\neq m\\
    \frac{\nu'_{1m}}{2a}\frac{1}{2}a^{2}\left[ J^{2}_{1}(\nu'_{1m})(1+1/\nu'^{2}_{1m})+J^{2}_{2}(\nu'_{1m})-J_{1}(\nu'_{1m})J_{3}(\nu'_{1m}) \right], &n=m
    \end{cases}\\
    &=&\delta_{mn}\frac{a}{2}(\nu'_{1m}-1/\nu'_{1m})J^{2}_{1}(\nu'_{1m}),\label{L4found2}
\end{eqnarray}
where the last equality was reached using the fact that $J_{3}(\nu'_{1m})=4 J_{2}(\nu'_{1m})/\nu'_{1m}-J_{1}(\nu'_{1m})$ and that $J_{0}(\nu'_{1m})=J_{2}(\nu'_{1m})=J_{1}(\nu'_{1m})/\nu'_{1m}$.
\end{widetext}



\bibliography{myrefs}

\end{document}